\documentclass[journal,a4paper,draftclsnofoot,onecolumn]{IEEEtran}
\usepackage{cite}
\usepackage{graphicx}
\usepackage{amssymb}
\usepackage{mathrsfs}
\usepackage{times}
\usepackage{cite}
\begin{document}
%*******************************************************************************
%*******************************************************************************
\thispagestyle{empty}
\begin{center}
{\LARGE Design and Development of a Burst Acquisition System for
     Geosynchronous Satcom Channels -- Final Report (2009-2010)}\\
     \vspace*{3in}
{\large {\it Project Investigator: K. Vasudevan\\
             Associate Professor\\
             email: vasu@iitk.ac.in}}\\
     
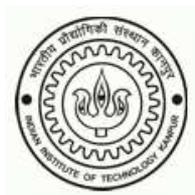
\vspace*{2in}
%*******************************************************************************
\begin{figure}[tbh]
\centering
\input{iitklogo.pstex_t}
\end{figure}
%*******************************************************************************
{Department of Electrical Engineering\\
 Indian Institute of Technology\\
 Kanpur - 208 016\\
 INDIA\\
October 2015}
\end{center}
%*******************************************************************************

%*******************************************************************************
%*******************************************************************************
\title{Design and Development of a Burst Acquisition System for
       Geosynchronous Satcom Channels -- Final Report (2009-2010)}
\author{Project Investigator: K. Vasudevan
\thanks{This project is supported by Defence Electronics Applications Lab
        (DEAL), Dehradun (ref. no. DEAL/02/4043/2005-2006/02/016).
	The project investigator is with the Dept. of Electrical Engg.,
        Indian Institute of Technology, Kanpur. Email: vasu@iitk.ac.in}}
%*******************************************************************************
\maketitle
%*******************************************************************************
\thispagestyle{empty}
%*******************************************************************************
%*******************************************************************************
\begin{abstract}
The key contribution of this work is to develop transmitter and receiver
algorithms in discrete-time for turbo-coded offset QPSK signals. The
proposed synchronization and detection techniques perform
effectively at an SNR per bit close to 1.5 dB, in the presence of a frequency
offset as large as 30 \% of the symbol-rate and a clock offset of 25 ppm (parts
per million).
Due to the use of up-sampling and matched filtering and a feedforward
approach, the
acquisition time for clock recovery is just equal to the length of the
preamble. The carrier recovery algorithm does not exhibit any phase ambiguity,
alleviating the need for differentially encoding the data at the transmitter.
The proposed techniques are well suited for discrete-time implementation.
\end{abstract}
%*******************************************************************************
%*******************************************************************************
\IEEEpeerreviewmaketitle
%*******************************************************************************
%*******************************************************************************
\section{Introduction}
\label{Sec:Intro}
Geosynchronous satellites provide line-of-sight communications with the
ground stations. Such communication links offer distortionless transmission,
with the only impairment being AWGN. Whereas transmit power is not so much
of an issue at the ground station, it is a precious commodity on-board the
satellite. With the growing demand for satellite
broadcast services, it has become necessary for the end users to receive
signals directly from the satellite. This calls for a vastly reduced
size and cost of the receiving equipment at the ground station (which is
usually the users premises or the handset) and superior modulation, coding
and synchronization techniques.  With the discovery of turbo codes, the
aforementioned scenario has become a reality.

In order to further improve the performance and bring down the cost of the
receivers, we propose offset QPSK as the modulation technique, which allows
the use of power efficient non-linear amplifiers and data-aided synchronization
algorithms which have a faster acquisition time than the non-data-aided
counterparts proposed in \cite{Amico01}. We also use the upsampled version
of the matched filter as an interpolator \cite{Harris01}, enabling the
implementation of a feedforward timing acquisition method, that is
faster than the feedback approach discussed in \cite{Amico01}. Though there
are other interpolation techniques \cite{Gardner93_1,Gardner93_2} for
timing acquisition, we believe that they are well suited for feedback
methods.

Before we proceed, a brief review of the literature on carrier and timing
synchronization is in order.
An improved phase lock loop that operates effectively at medium to low
SNR is proposed in \cite{Daley74}. A data-aided carrier recovery loop for
duobinary encoded offset QPSK is given in \cite{Taylor79}.
A tutorial on carrier and timing synchronization is given in \cite{Franks80}.
Joint carrier recovery and equalization of digitally modulated signals is
discussed in \cite{Chang80,Godard80}.
A data-aided carrier recovery algorithm for estimating phase and frequency
offsets is discussed in \cite{Cupo89,Nestor99} and also in
\cite{Saito89,Kato92_1} for digital land mobile radio and satellite
communication. Detection of bursty QPSK signals at low SNR is described in
\cite{Miyo91}. A digital modem for offset-QPSK is dealt with in \cite{Seki92}.
Carrier synchronization for trellis-coded signals is given in
\cite{Pent89}. A digital PLL
for QPSK signals is described in \cite{Xezonakis93} and an all-digital
implementation of carrier synchronization for digital radio systems is
proposed in \cite{Sari91}. A comparison of different carrier recovery
techniques is presented in \cite{Benani00}. A carrier recovery algorithm
for $M$-ary QAM with a capability to track large frequency offsets is discussed
in \cite{Choi01}.
A non-data-aided carrier recovery method for modified 128-QAM
is proposed in \cite{Hou05}.

Timing recovery can be broadly classified into synchronous and asynchronous
methods. In the synchronous methods, the local clock at the receiver is
regenerated from the incoming signal. Such techniques are implemented in
hardware and are usually employed in analog modems \cite{Feher78}. In the
asynchronous approach, the local clock at the receiver is free running, due to
which the ratio of the receiver sampling frequency to the incoming symbol-rate
is not an
integer. If this ratio were to be an integer (equal to say, $M$), then timing
acquisition alone would have sufficed -- we could recover the symbols from
the matched filter output every $M^{th}$ sample. Let us denote the sampling
epoch as $m_0$ modulo-$M$. However, in practical
situations the transmitter and receiver clocks are asynchronous (their
frequencies are not exactly identical), therefore
the above mentioned ratio is not an integer. Hence timing needs to be
acquired and tracked (this is explained in the next para and in
section~\ref{SSec:Sync_Track}). In the present
context, the word ``timing'' implies
knowing when to recover the symbols from the matched filter output. In the
asynchronous methods of timing recovery, signal processing techniques are used
which are suitable for software implementation
\cite{Koblents92,Serpedin04,Shirato05}.

The transmit and receive clocks are usually specified as $F_0$ Hz,
$\pm \delta$ parts per million (ppm).
This implies that the actual frequency of the transmit and receive clocks lies
in the range $F_0(1\pm\delta\times 10^{-6})$ Hz. The worst case frequency
difference is $\pm 2F_0\delta\times 10^{-6}$ Hz. For ease of understanding,
we could assume that the transmit clock is exact ($F_0$ Hz) and the
receive clock is $F_0$ Hz, $\pm 2\delta$ ppm. Let us now assume that the
transmitted signal is sampled using the exact clock (with frequency $F_0$
Hz) and $10^6$ samples are obtained over a period of time. If the same
transmitted signal is sampled using the receive clock (having an accuracy of
$\pm 2\delta $ ppm) over the same time interval, we would obtain
$10^6 \pm 2\delta$ samples. For example, if $\delta=0.5$, then we would
obtain one sample more or less over $10^6$ samples, which further implies that
the sampling epoch would change by one sample over $10^6$ samples (the new
sampling epoch would be $(m_0 \pm 1)$ modulo-$M$).

More recently, iterative timing recovery is proposed in \cite{Nayak04}.
The Cram{\'e}r-Rao bound for non-data-aided timing recovery for linearly
modulated signals with no ISI is presented in \cite{Bergel03}.

This paper is organized as follows. Section~\ref{Sec:Sys_Model} discusses
the system model. In Section~\ref{Sec:Receiver} we discuss the
receiver algorithms. The performance results are discussed in
Section~\ref{Sec:Results}. Finally, in Section~\ref{Sec:Conclude} we present
our conclusions.

%*******************************************************************************
\section{System Model}
\label{Sec:Sys_Model}
%*******************************************************************************
\begin{figure}[tbh]
\centering
\input{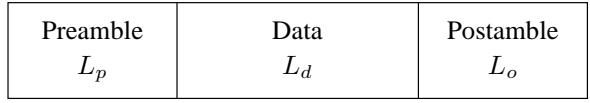}
\caption{The burst structure.}
\label{Fig:Burst}
\end{figure}
%*******************************************************************************
We assume that the data to be transmitted is organized into bursts (or frames).
The burst consists of a known preamble of length $L_p$ (QPSK) symbols,
followed by turbo-coded data of length $L_d$ symbols and a known postamble of
length $L_o$ symbols. Thus, the total length of the frame is
%*******************************************************************************
\begin{eqnarray}
\label{Eq:Pap7_Eq0_1}
L = L_p + L_d + L_o.
\end{eqnarray}
%*******************************************************************************
The $L_d$ QPSK symbols are obtained by passing an uncoded bit stream through a
rate-$1/2$ turbo code. The generator matrix of each of the constituent
encoders is given by \cite{Ryan99}:
%*******************************************************************************
\begin{eqnarray}
\label{Eq:Pap7_Eq0_2}
\mathbf{G}(D) =
\left[
\begin{array}{cc}
 1 & \frac{\displaystyle 1+D^2+D^3+D^4}{\displaystyle 1+D+D^4}
\end{array}
\right].
\end{eqnarray}
%*******************************************************************************
The received signal can be written as:
%*******************************************************************************
\begin{eqnarray}
\label{Eq:Pap7_Eq1}
r(t) = \sqrt{4/T}
       \,
       \Re
       \left
       \{
       \tilde s(t)
       \mathrm{e}^{\,\mathrm{j}\, [2\pi (F_c + \Delta F)t+\theta_0]}
       \right
       \} + w_1(t)
\end{eqnarray}
%*******************************************************************************
where $\Re\{\cdot\}$ denotes the real-part, $1/T$ is the baud-rate,
$F_c$ is the nominal carrier frequency, $\Delta F$ is the frequency
offset (which can be positive or negative), $\theta_0$ is the carrier phase
and $w_1(t)$ is additive white Gaussian noise with two-sided power spectral
density $N_0/2$ (watts/Hz). The term $\tilde s(t)$ (we use tilde to
denote complex quantities) in (\ref{Eq:Pap7_Eq1})
is the complex envelope of the offset QPSK signal and is given by:
%*******************************************************************************
\begin{eqnarray}
\label{Eq:Pap7_Eq2}
\tilde s(t) & = &  s_I(t) + \mathrm{j}\, s_Q(t)                \nonumber  \\
            & = & \sum_{k=0}^{L-1} S_{k,\, I} p(t-kT-\alpha)
                  \nonumber  \\
            &   & \mbox{ } +
                  \mathrm{j}\,
                  \sum_{k=0}^{L-1} S_{k,\, Q} p(t-kT-T/2-\alpha)
\end{eqnarray}
%*******************************************************************************
where $S_{k,\, I}\in \pm 1$ and $S_{k,\, Q}\in \pm 1$ are the in-phase and
quadrature components of the QPSK symbol and
$p(t)$ is the impulse response of the transmit filter, which is
assumed to have the root-raised cosine spectrum with $40\%$ roll-off. The
variable $\alpha$ denotes the random timing phase which is assumed to be
uniformly distributed in $[0,\, T)$. Observe that $p(t)$ extends over
$-\infty < t < \infty$. However in practice, it can be delayed and truncated
to obtain
a causal and finite impulse response, with negligible intersymbol-interference
(ISI) at the matched filter output. The task of the receiver is to
estimate the QPSK symbols such that the error-rate is close to the theoretical
limit.

Since we deal with discrete-time signals in this paper, the first task is to
convert $r(t)$ in (\ref{Eq:Pap7_Eq1}) into a discrete-time signal. This is
accomplished by passing $r(t)$ through a bandpass filter (BPF) followed by
bandpass sampling. For convenience of subsequent analysis, we assume an
ideal BPF having unit energy with a gain of $\sqrt{T/4}$ in the passband
extending over $[F_c-1/T,\, F_c + 1/T]$ Hz.
This ensures that the noise power at the BPF output is $N_0/2$. Assuming
a sampling frequency of $F_s=1/T_s$, the bandpass sampling requirements can be
stated as follows \cite{Vasu_Book}:
%*******************************************************************************
\begin{eqnarray}
\label{Eq:Pap7_Eq3}
\frac{2\pi (F_c-1/T)}{F_s} & \ge & k\pi                \nonumber  \\
\frac{2\pi (F_c+1/T)}{F_s} & \le & (k+1)\pi
\end{eqnarray}
%*******************************************************************************
where $k$ is a positive integer. We assume that the conditions in
(\ref{Eq:Pap7_Eq3}) are satisfied with an equality so that $F_s=4/T$ or
equivalently $T/T_s=4$. Therefore, the symbols can be delayed by $T/2 \equiv 2$
samples in discrete-time. We further assume that
%*******************************************************************************
\begin{eqnarray}
\label{Eq:Pap7_Eq4}
\frac{2\pi F_c}{F_s} = k\pi + \pi/2 = \pi/2 \bmod 2\pi
\end{eqnarray}
%*******************************************************************************
if $k$ is even. Thus the
output of the analog-to-digital (A/D) converter after bandpass sampling
can be written as
%*******************************************************************************
\begin{eqnarray}
\label{Eq:Pap7_Eq5}
r(nT_s) = \Re
          \left
          \{
          \tilde s(nT_s)
          \mathrm{e}^{\,\mathrm{j}\, [(\pi/2 + \omega_0)n+\theta_0]}
          \right
          \} + w(nT_s)
\end{eqnarray}
%*******************************************************************************
where
%*******************************************************************************
\begin{eqnarray}
\label{Eq:Pap7_Eq6}
\tilde s(nT_s) & = & \tilde s(t)|_{t=nT_s}           \nonumber  \\
               & = &  s_I(nT_s) + \mathrm{j} \,
                      s_Q(nT_s)                      \nonumber  \\
w(nT_s)        & = &  w(t)|_{t=nT_s}                 \nonumber  \\
\omega_0       & = & 2\pi \Delta F/F_s
\end{eqnarray}
%*******************************************************************************
where $w(t)$ is noise at the BPF output.
Note that $w(nT_s)$ denotes samples white Gaussian noise with variance
$N_0/2$.
%*******************************************************************************
\begin{figure}[tbh]
\centering
\input{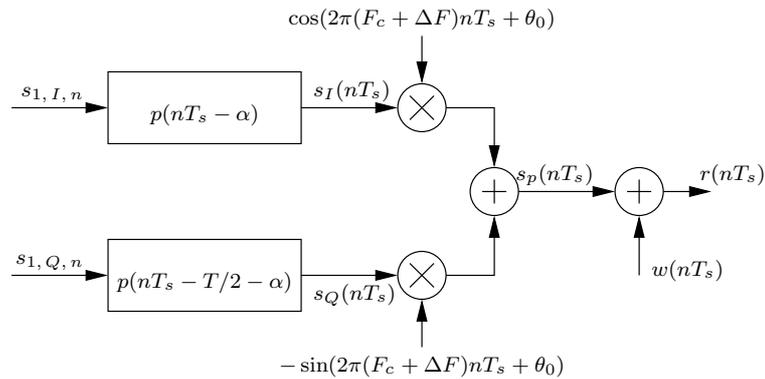}
\caption{Simulation model of the transmitter.}
\label{Fig:Pap8_Lin_Tx}
\end{figure}
%*******************************************************************************
%*******************************************************************************
\begin{figure}[tbh]
\centering
\input{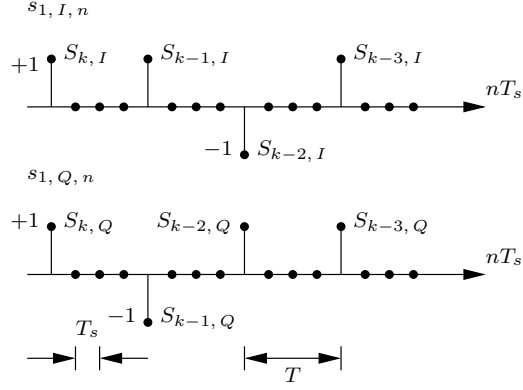}
\caption{Symbol sequence input to the transmitter. Here $T/T_s=M=4$.}
\label{Fig:Pap8_Delta1}
\end{figure}
%*******************************************************************************
The simulation model of the transmitter and its input are illustrated in
Figures~\ref{Fig:Pap8_Lin_Tx} and \ref{Fig:Pap8_Delta1}.

Having assumed that $T/T_s=4$, it is now necessary to find out what is the
maximum frequency offset that can be tolerated. With $40\%$ roll-off,
the bandwidth of the complex baseband signal is $1.4/(2T)=0.7/T$.
%*******************************************************************************
\begin{figure}[tbh]
\centering
\input{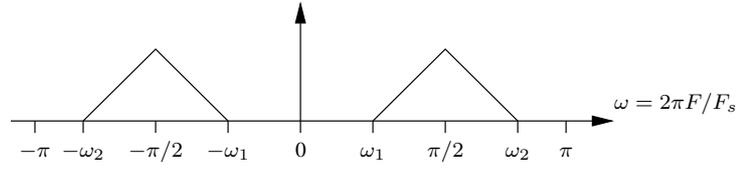}
\caption{The spectrum of the discrete-time sampled signal with $\Delta F=0$.}
\label{Fig:Foff_Spec}
\end{figure}
%*******************************************************************************
The spectrum of the discrete-time sampled signal is shown in
Figure~\ref{Fig:Foff_Spec}, when the frequency-offset $\Delta F = 0$.
Here
%*******************************************************************************
\begin{eqnarray}
\label{Eq:Pap7_Eq7}
\omega_1 & = & \pi/2 - 2\pi \times 0.7/(TF_s) = 0.15\pi     \nonumber  \\
\omega_2 & = & \pi/2 + 2\pi \times 0.7/(TF_s) = 0.85\pi = \pi - \omega_1.
\end{eqnarray}
%*******************************************************************************
In order to avoid aliasing of the spectrum, we require
%*******************************************************************************
\begin{eqnarray}
\label{Eq:Pap7_Eq8}
|2\pi \Delta F/F_s| & \le & \omega_1         \nonumber  \\
\Rightarrow
|\omega_0|          & \le & \omega_1.
\end{eqnarray}
%*******************************************************************************
Substituting for $\omega_1$ from (\ref{Eq:Pap7_Eq7}) and using $T/T_s=4$, we
obtain
%*******************************************************************************
\begin{eqnarray}
\label{Eq:Pap7_Eq8_1}
|\Delta F| \le 0.3/T
\end{eqnarray}
%*******************************************************************************
which is less than or equal to 30\% of the symbol-rate. Having justified the
maximum frequency offset used in this work, we next discuss the procedure for
simulating the clock offset.

Assume that $r(t)$ in (\ref{Eq:Pap7_Eq1}) is sampled at a rate of
%*******************************************************************************
\begin{eqnarray}
\label{Eq:Pap7_Eq9}
F_s' = F_s(1 \pm 2 \delta \times 10^{-6}).
\end{eqnarray}
%*******************************************************************************
The corresponding sampling period is:
%*******************************************************************************
\begin{eqnarray}
\label{Eq:Pap7_Eq10}
T_s' = 1/F_s' = T_s(1 \mp 2 \delta \times 10^{-6})
\stackrel{\Delta}{=} T_s + \epsilon.
\end{eqnarray}
%*******************************************************************************
Then
%*******************************************************************************
\begin{eqnarray}
\label{Eq:Pap7_Eq11}
r(nT_s') = \Re
           \left
           \{
           \tilde s(nT_s')
           \mathrm{e}^{\,\mathrm{j}\, [2\pi (F_c + \Delta F)nT_s'+\theta_0]}
           \right
           \} + w_1(nT_s').
\end{eqnarray}
%*******************************************************************************
Define
%*******************************************************************************
\begin{eqnarray}
\label{Eq:Pap7_Eq12}
\omega_3   =    2
               \pi
               \frac{\Delta F \mp (F_c + \Delta F)2\delta \times 10^{-6}}
                    {F_s}.
\end{eqnarray}
%*******************************************************************************
Then
%*******************************************************************************
\begin{eqnarray}
\label{Eq:Pap7_Eq13}
r(nT_s') = \Re
           \left
           \{
           \tilde s(nT_s')
           \mathrm{e}^{\,\mathrm{j}\, [(\pi/2 + \omega_3)n+\theta_0]}
           \right
           \} + w_1(nT_s').
\end{eqnarray}
%*******************************************************************************
Thus, one of the effects of having an error in the sampling frequency is to
introduce an additional frequency offset equal to:
%*******************************************************************************
\begin{eqnarray}
\label{Eq:Pap7_Eq14}
\mp 2\pi \frac{(F_c + \Delta F)2\delta \times 10^{-6}}{F_s}
\qquad \mbox{radians}.
\end{eqnarray}
%*******************************************************************************
Henceforth, we assume that $|\omega_3| \le \omega_1$ to avoid aliasing (see
also (\ref{Eq:Pap7_Eq8})). We now
study the effect of $T_s'$ on the complex baseband signal $\tilde s(\cdot)$ in
(\ref{Eq:Pap7_Eq13}).

We have
%*******************************************************************************
\begin{eqnarray}
\label{Eq:Pap7_Eq15}
\tilde s(nT_s') & = & \sum_{k=0}^{L-1}
                       S_{k,\, I} p(nT_s'-kT-\alpha)
                      \nonumber  \\
                &   & \mbox{ } +
                      \mathrm{j}\,
                      \sum_{k=0}^{L-1}
                       S_{k,\, Q} p(nT_s'-kT-T/2-\alpha)
\end{eqnarray}
%*******************************************************************************
Let the impulse response of the transmit filter extend over
$[0,\, N]$ samples, at a
sampling rate of $F_s=4/T$. Let
%*******************************************************************************
\begin{eqnarray}
\label{Eq:Pap7_Eq15_1}
nT_s - kT = iT_s
\end{eqnarray}
%*******************************************************************************
where $i$ is an integer. Using (\ref{Eq:Pap7_Eq10}), (\ref{Eq:Pap7_Eq15})
can be rewritten as:
%*******************************************************************************
\begin{eqnarray}
\label{Eq:Pap7_Eq16}
\tilde s(nT_s') & = & \sum_{k=0}^{L-1}
                       S_{k,\, I} p(nT_s+n\epsilon-kT-\alpha)
                      \nonumber  \\
                &   & \mbox{ } +
                      \mathrm{j}\,
                      \sum_{k=0}^{L-1}
                       S_{k,\, Q} p(nT_s+n\epsilon-kT-T/2-\alpha)
                      \nonumber  \\
                & = & \sum_{i=0}^{N}
                       p(iT_s+n\epsilon-\alpha) S_{(n-i)/4,\, I}
                      \nonumber  \\
                &   & \mbox{ } +
                      \mathrm{j}\,
                      \sum_{i=0}^{N}
                       p(iT_s+n\epsilon-T/2-\alpha) S_{(n-i)/4,\, Q}
\end{eqnarray}
%*******************************************************************************
where it is understood that $S_{(n-i)/4,\, I}$ and $S_{(n-i)/4,\, Q}$ are
non-zero every $4^{th}$ value of $n$, all other values are zero (see
Figure~\ref{Fig:Pap8_Delta1}).

\begin{figure}[tbh]
\centering
\input{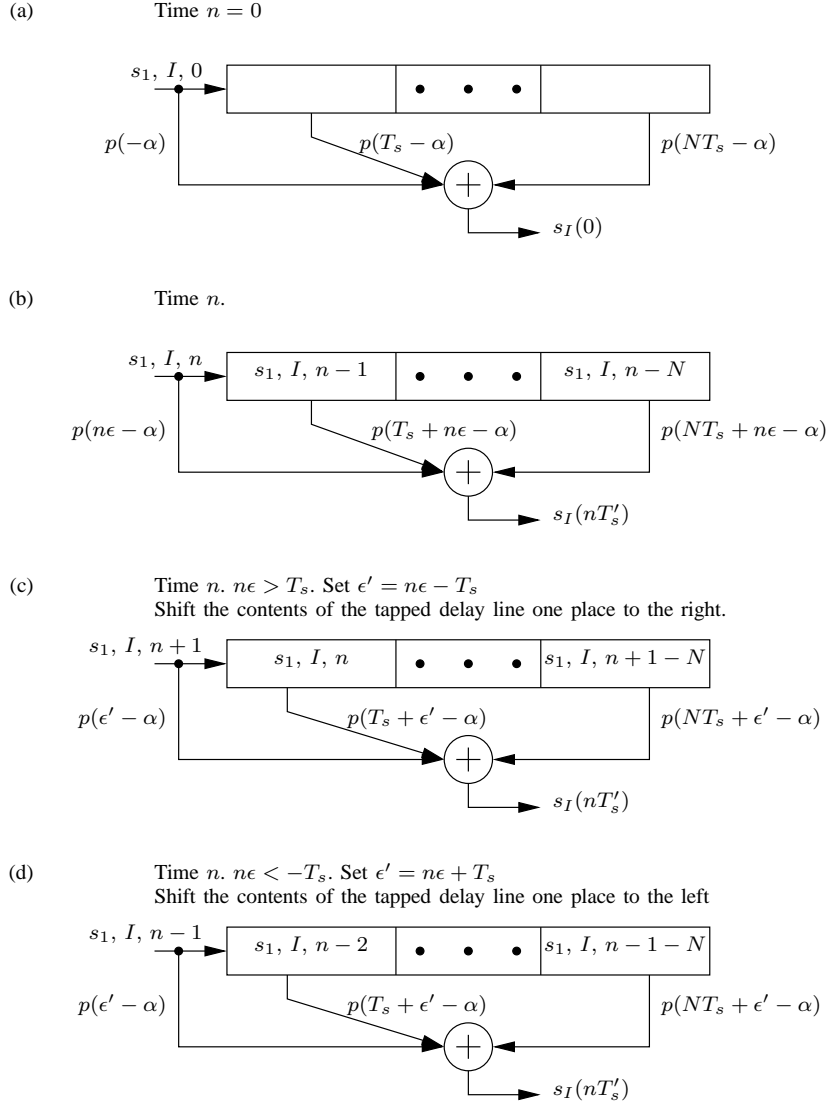}
\caption{Illustrating the process of generating $s_I(nT_s')$ (a) Time $n=0$.
(b) Time $n$. (c) Adjustments at time $n$ when $n\epsilon > T_s$.
(d) Adjustments at time $n$ when $n\epsilon < -T_s$.}
\label{Fig:Tx_Flt_Clk_Err}
\end{figure}

The above equation suggests that the complex baseband signal must be
generated using time-varying transmit filter coefficients \cite{Vasu_MS}.
Note that if
$\epsilon=0$, the transmit filter would be time-invariant, since $\alpha$
is a constant for a given burst. Clearly, it doesn't make sense to allow
$n\epsilon$ to grow without bound. In fact, it needs to be periodically
``normalized''. How this is done, is shown in
Figure~\ref{Fig:Tx_Flt_Clk_Err}, for generating the in-phase part of the
complex baseband \cite{Vasu_MS}. The sequence $s_{1,\, I,\, n}$ is illustrated
in Figure~\ref{Fig:Pap8_Delta1}. The procedure for generating the quadrature
part of the complex baseband is similar. Note that the signal $s_I(nT_s')$
obtained from Figure~\ref{Fig:Tx_Flt_Clk_Err}(b) is approximately equal to
that obtained in parts (c) and (d), as long as the first and the last transmit
filter coefficients contribute insignificantly to the energy of the overall
transmit filter. This can be ensured by having a large enough $N$.
The simulation model for the transmitter is similar to
Figure~\ref{Fig:Pap8_Lin_Tx} with $T_s$ replaced by $T_s'$. Since
$T_s' \approx T_s$, we continue to assume that the samples of $w(nT_s')$ are
uncorrelated with zero mean and variance $N_0/2$. The next step is to discuss
the receiver.
%*******************************************************************************
%*******************************************************************************
%*******************************************************************************
\section{Receiver}
\label{Sec:Receiver}
%*******************************************************************************
\begin{figure}[tbh]
\centering
\input{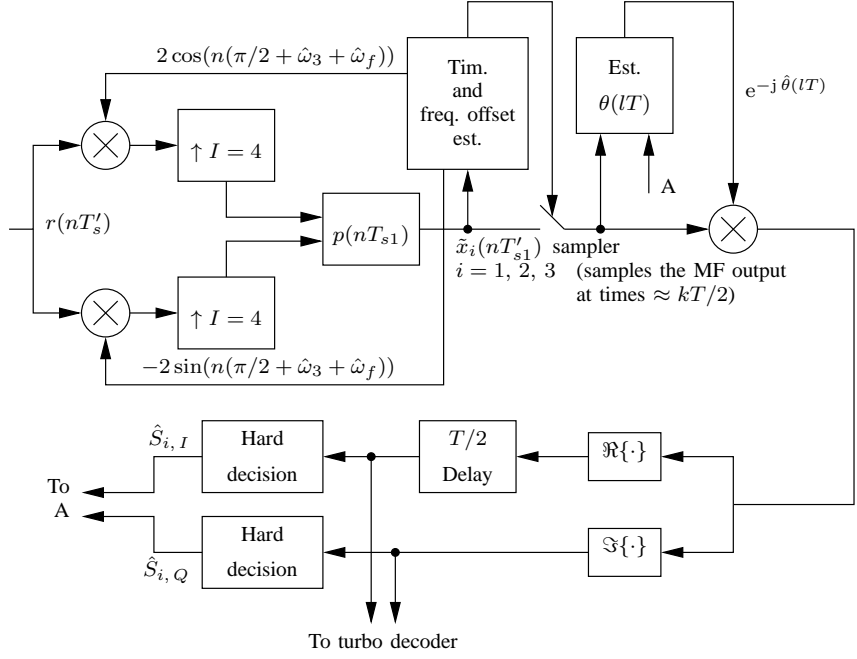}
\caption{The discrete-time receiver which up-samples the local
         oscillator output by a factor of $I$ and then performs matched
         filtering at a sampling frequency $F_{s1}$.}
\label{Fig:Rx1}
\end{figure}
%*******************************************************************************
Let $\hat\omega_3$ and $\hat\omega_f$ denote the ``coarse'' and the ``fine''
estimates of the frequency offset respectively, corresponding to the
sampling frequency of $1/T_s$.
The receiver in Figure~\ref{Fig:Rx1} operates in the following steps:
%*******************************************************************************
\begin{enumerate}
 \item Demodulate $r(nT_s')$ with $\hat\omega_3=\hat\omega_f=0$ in
       Figure~\ref{Fig:Rx1} and detect the start of frame using the
       ``differential'' correlation method. Store the samples $r(nT_s')$
       corresponding to a frame. Index the first incoming sample as 0,
       the next sample as 1 and so on.
       Obtain a coarse estimate of the frequency offset and denote it
       as $\hat\omega_3$.
 \item Demodulate the stored values of $r(nT_s')$ using $\pi/2+\hat\omega_3$.
       Estimate the start of frame for the second time using the differential
       correlation method.
 \item Obtain the maximum likelihood estimate of the
       residual frequency offset which is equal to
       $\omega_f=\omega_3-\hat\omega_3$. Denote this estimate as
       $\hat\omega_f$.
 \item Demodulate the stored values of $r(nT_s')$ using
       $\pi/2+\hat\omega_3+\hat\omega_f$.
       Estimate the start of frame for the third time using the correlation
       method. Estimate the symbol
       amplitude, $\theta_0$ and the variance of additive noise.
 \item Detect the data using the turbo decoder. Track the timing and
       carrier phase resulting due to the clock offset and residual frequency
       offset given by $\omega_f-\hat\omega_f$ respectively.
\end{enumerate}
%*******************************************************************************
Before we proceed to elaborate on the above steps, let us look into the
operation of up-sampling and matched filtering \cite{Vasu08}.
%*******************************************************************************
\subsection{Up-Sampling and Matched Filtering}
\label{SSec:Receiver}
%*******************************************************************************
%*******************************************************************************
\begin{figure}[tbh]
\centering
\input{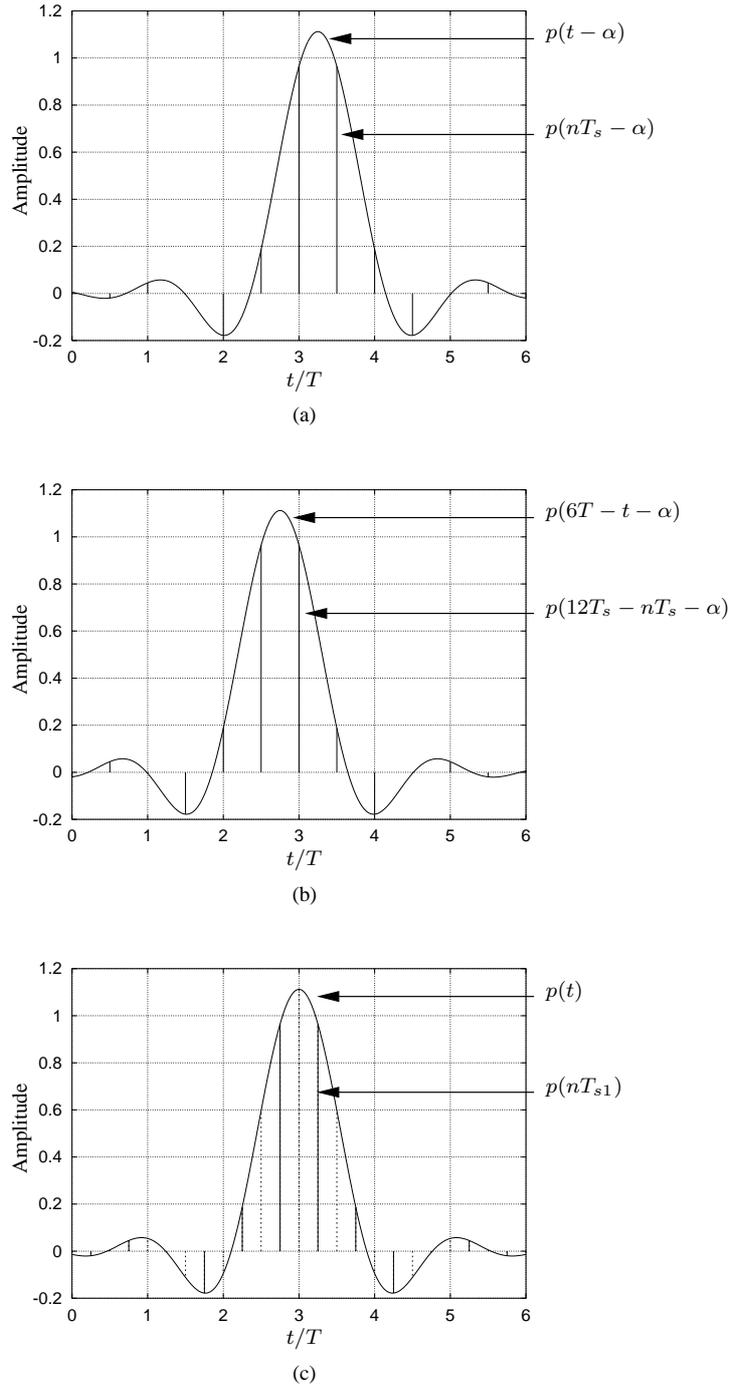}
\caption{(a) The received pulse $p(t-\alpha)$ and its samples taken at
$T/T_s=2$. $\alpha=T_s/4$. (b) Filter matched to the received pulse
$p(6T-t-\alpha)$ and its
samples taken at $T/T_s=2$. (c) Filter matched to $p(t)$ (in this case is
$p(6T-t)=p(t)$ itself) sampled at a higher frequency $T/T_{s1}=4$. Observe
that one
set of samples (shown by solid impulses) correspond to the samples of the
matched filter in part (b), for $\alpha=T_s/4$.}
\label{Fig:Interpol}
\end{figure}
%*******************************************************************************
For ease of exposition, we assume that $\delta=0$ ($T_s'=T_s$). The samples of
the transmitted pulse is shown in Figure~\ref{Fig:Interpol}(a) by solid
impulses (Kronecker delta function). We assume that $p(t-\alpha)$ spans over
$N+1(=13)$ ($0 \le n \le N$ ) samples and the sampling frequency $1/T_s$ is
such
that it satisfies the Nyquist criterion for no aliasing of the spectrum of
$p(t)$. The corresponding discrete-time matched filter is shown in
Figure~\ref{Fig:Interpol}(b) by solid impulses. The matched filter
output can be obtained using the frequency-domain approach. Let $P(F)$
denote the Fourier transform of $p(t)$, assumed to be bandlimited to
$|F|\le B$. Then the discrete-time Fourier transform of $p(nT_s-\alpha)$ is
obtained as follows:
%*******************************************************************************
\begin{eqnarray}
\label{Eq:Ch4_Tim_Sync_Eq2}
p(t-\alpha)    & = &  g(t)                                \nonumber  \\
               &     \rightleftharpoons
                   & \tilde G(F)                          \nonumber  \\
               & = & \left
                     \{
                     \begin{array}{ll}
                     \tilde P(F)
                     \mathrm{e}^{-\mathrm{j}\, 2\pi F\alpha} &
                     \mbox{$-B \le F \le B$}\\
                      0                                      &
                     \mbox{otherwise}
                     \end{array}
                     \right.                              \nonumber  \\
\end{eqnarray}
%*******************************************************************************
Therefore
%*******************************************************************************
\begin{eqnarray}
\label{Eq:Ch4_Tim_Sync_Eq2_1}
p(nT_s-\alpha) & = &  g(nT_s)                             \nonumber  \\
               &     \rightleftharpoons
                   & \tilde G_{\mathscr{P}}(F)            \nonumber  \\
               & = & \left
                     \{
                     \begin{array}{ll}
                     \tilde G(F)/T_s &
                     \mbox{$0 \le     |F| \le B$}\\
                      0                                      &
                     \mbox{$B \le |F| \le F_s/2$}
                     \end{array}
                     \right.
\end{eqnarray}
%*******************************************************************************
where $\tilde G_{\mathscr{P}}(F)$ denotes a periodic function of frequency,
that is \cite{Vasu_Book}
%*******************************************************************************
\begin{eqnarray}
\label{Eq:Ch4_Tim_Sync_Eq2_2}
\tilde G_{\mathscr{P}}(F) =
\tilde G_{\mathscr{P}}(F+kF_s) = \frac{1}{T_s}
                                 \sum_{i=-\infty}^{\infty}
                                 \tilde G(F - iF_s).
\end{eqnarray}
%*******************************************************************************
Now, when $p(nT_s-\alpha)$ is convolved with
$p(NT_s-\alpha-nT_s)$, the peak occurs at $t=NT_s$, independent
of $\alpha$. Note that $p(NT_s-\alpha-nT_s)$ represents the discrete-time
causal matched filter for $p(nT_s-\alpha)$. The peak value is equal to
$R_{pp}(0)/T_s$ \cite{Vasu_Book}, where $R_{pp}(t)$
is the continuous-time autocorrelation of $p(t)$.
In practice however since the receiver does not know $\alpha$,
it is not possible to obtain the exact matched filter as in
Figure~\ref{Fig:Interpol}(b).
The solution lies in sampling
$p(NT_s-t)$ at a higher frequency (say $F_{s1}=1/T_{s1}$) compared to $1/T_s$,
as illustrated in Figure~\ref{Fig:Interpol}(c).

Let us denote $T_s/T_{s1} = I$ which is referred to as the
interpolation factor.
Construct the up-sampled sequence from the incoming signal $p(nT_s-\alpha)$
as follows:
%*******************************************************************************
\begin{eqnarray}
\label{Eq:Ch4_Tim_Sync_Eq3}
g_1(nT_{s1}) =
\left
\{
\begin{array}{ll}
g(nT_s/I) & \mbox{for $n=mI$}\\
0         & \mbox{otherwise}
\end{array}
\right.
\end{eqnarray}
%*******************************************************************************
where $g(nT_s)$ is defined in (\ref{Eq:Ch4_Tim_Sync_Eq2_1}).
The discrete-time Fourier transform (DTFT) of $g_1(nT_{s1})$ is
\cite{Proakis92}:
%*******************************************************************************
\begin{eqnarray}
\label{Eq:Ch4_Tim_Sync_Eq4}
g_1(nT_{s1}) \rightleftharpoons G_{\mathscr{P}}(FI).
\end{eqnarray}
%*******************************************************************************
Let us define a new frequency variable $F_1=FI$
with respect to the new sampling frequency $F_{s1}$.
Now, if $p(NT_s-t)$ is sampled at a rate $F_{s1}$ the DTFT of the resulting
sequence is:
%*******************************************************************************
\begin{eqnarray}
\label{Eq:Ch4_Tim_Sync_Eq5}
p(NT_s-nT_{s1}) =  g_2(nT_{s1}) \rightleftharpoons &
                  \tilde G_{\mathscr{P},\, 2}(F_1)
\end{eqnarray}
%*******************************************************************************
where
%*******************************************************************************
\begin{eqnarray}
\label{Eq:Ch4_Tim_Sync_Eq5_1}
\lefteqn{
\tilde G_{\mathscr{P},\, 2}(F_1)}    \nonumber  \\
& = &
\left
\{
\begin{array}{ll}
\frac{\displaystyle \tilde P^*(F_1)}{\displaystyle T_{s1}}
\,\,\mathrm{e}^{-\mathrm{j}\, 2\pi F_1 N T_s} &
\mbox{$0 \le  |F_1| \le B$}\\
0                                                                &
\mbox{$B \le |F_1| \le F_{s1}/2$}.
\end{array}
\right.                              \nonumber  \\
\end{eqnarray}
%*******************************************************************************
The convolution of $g_1(nT_{s1})$ with $g_2(nT_{s1})$ can be written as
\cite{Vasu_Book}:
%*******************************************************************************
\begin{eqnarray}
\label{Eq:Ch4_Tim_Sync_Eq6}
\lefteqn{
g_1(nT_{s1}) \star g_2(nT_{s1})}                          \nonumber  \\
                           & = & \frac{1}{T_s T_{s1} F_{s1}}
                                 \int_{F_1=-B}^{B}
                                 \left|
                                 \tilde P(F_1)
                                 \right|^2                \nonumber  \\
                           &   & \mbox{ } \times
                                 \mathrm{e}^{\,\mathrm{j}\, 2\pi F_1
                                            (nT_{s1}-\alpha-NIT_{s1})}\, dF_1.
\end{eqnarray}
%*******************************************************************************
Clearly if $\alpha=n_0T_{s1}$, where $n_0$ is an integer, the above
convolution becomes
%*******************************************************************************
\begin{eqnarray}
\label{Eq:Ch4_Tim_Sync_Eq7}
g_1(nT_{s1}) \star g_2(nT_{s1}) = \frac{R_{pp}((n-n_0-NI)T_{s1})}{T_s}
\end{eqnarray}
%*******************************************************************************
with a peak value equal to $R_{pp}(0)/T_s$, occurring at $(n_0+NI)T_{s1}$.
When $\alpha$ is not an integer multiple of $T_{s1}$, $R_{pp}(0)$ occurs
in between two consecutive samples and we can get close to the peak
by increasing $I$.
Henceforth, we assume that the first symbol in the frame occurs
at time $\alpha+NIT_{s1}$ and $R_{pp}(0)/T_s=1$. If the clock offset $\delta=0$
then the subsequent
symbols can be extracted from the matched filter output at times
$(n_0+NI)T_{s1}+kT/2=(n_0+NI)T_{s1}+2kIT_{s1}$ where we have used the fact that
%*******************************************************************************
\begin{eqnarray}
\label{Eq:Ch4_Tim_Sync_Eq8}
\frac{T}{T_{s1}} = \frac{T}{T_s} \cdot \frac{T_s}{T_{s1}} = 4I.
\end{eqnarray}
%*******************************************************************************
If $\delta \ne 0$, then the sampling instants at the MF output varies with
time and needs to be tracked. This aspect will be taken up in
subsection~\ref{SSec:Sync_Track}.
However, the matched filter coefficients are obtained at $F_{s1}$ and not
at $F_{s1}'$, since in practice $F_{s1}'$ is not known at the receiver.

The derivation of the signal at the input and the output of the matched filter
in the presence of a frequency offset and clock offset
is too involved. We are only interested in the signal at the output of the
sampler. This will be taken up in the next subsection.

%*******************************************************************************
\subsection{Timing and Carrier Acquisition using the Preamble}
\label{SSec:Sync_Acq}
%*******************************************************************************
This section is an elaboration of the first four steps listed in
section~\ref{Sec:Receiver}. Recall that in the first step, the start of frame
and a coarse estimate of the frequency offset is obtained. The demodulation
is done using a local oscillator frequency of $\pi/2$ radians. Let
%*******************************************************************************
\begin{eqnarray}
\label{Eq:Sync_Eq0}
t_0=\alpha+NIT_{s1}\approx \alpha+NIT_{s1}'.
\end{eqnarray}
%*******************************************************************************
We proceed by making a key observation that at the right instants,
the $T$-spaced sampler output can be approximated as (since $R_{pp}(0)/T_s=1$):
%*******************************************************************************
\begin{eqnarray}
\label{Eq:Sync_Eq1}
\tilde x_1(t_0+kT) \approx
                   \tilde \beta(kT)
                   \mathrm{e}^{\,\mathrm{j}\,(\omega_3 M k+\theta_0)} +
                   \tilde v_1(t_0+kT)
\end{eqnarray}
%*******************************************************************************
for $0\le k \le L-1$ where
%*******************************************************************************
\begin{eqnarray}
\label{Eq:Sync_Eq2}
\tilde \beta(kT) =  S_{k,\, I} + \mathrm{j}\, \gamma_{k,\, Q}
\end{eqnarray}
%*******************************************************************************
where $S_{k,\, I}$ is defined in (\ref{Eq:Pap7_Eq2}) and $\gamma_{k,\, Q}$
denotes the intersymbol interference (ISI) in the quadrature arm.
%*******************************************************************************
\begin{figure}[tbh]
\centering
\input{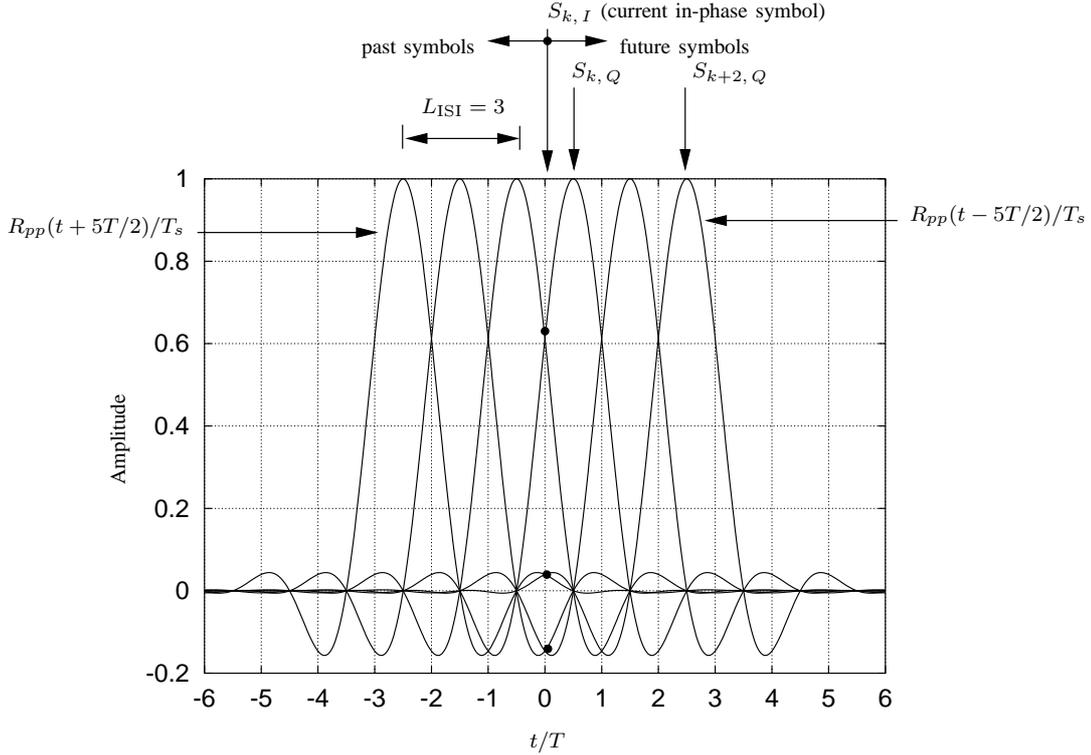}
\caption{Procedure for obtaining $h_j$. Signal in the quadrature arm is
         $(1/T_s)\sum_k S_{k,\, Q} R_{pp}(t-kT-T/2)$. From the dots at time
         $t=0$ we get $h_0=0.041$, $h_1=-0.147$, $h_2=0.612$, $h_3=0.612$,
         $h_4=-0.147$ and $h_5=0.041$.}
\label{Fig:ISI}
\end{figure}
%*******************************************************************************
Note that, $\gamma_{k,\, Q}$ is a function of the past, current
and future quadrature symbols as given by (see Figure~\ref{Fig:ISI}):
%*******************************************************************************
\begin{eqnarray}
\label{Eq:Sync_Eq5_1}
\gamma_{k,\, Q} = \sum_{j=0}^{2L_{\mathrm{ISI}}-1}
                   S_{k+L_{\mathrm{ISI}}-1-j,\, Q} h_j
\end{eqnarray}
%*******************************************************************************
where $L_{\mathrm{ISI}}$ denotes the span of $R_{pp}(t)$ (in symbol durations)
on the either side of $R_{pp}(0)$ and $h_j$ denotes coefficients of the filter
having a raised-cosine frequency response. In the simulations,
$L_{\mathrm{ISI}}$ was taken to be 3.

The term $\tilde v(\cdot)$ in (\ref{Eq:Sync_Eq1}) denotes samples of zero-mean
Gaussian noise with autocorrelation \cite{Vasu_Book}
%*******************************************************************************
\begin{eqnarray}
\label{Eq:Sync_Eq3}
(1/2) E
\left[
\tilde v_1(kT)
\tilde v_1^*(kT-mT)
\right] = N_0 \delta_K(mT)
\end{eqnarray}
%*******************************************************************************
where $\delta_K(\cdot)$ is the Kronecker delta function. The approximation in
(\ref{Eq:Sync_Eq1}) is due to the presence of ISI, besides noise. As
$\omega_3$ and $\delta$ tend to zero, and when
$t_0$ is an integer multiple of $T_{s1}$, the ISI approaches zero, and
the approximation becomes an equality.

Define
%*******************************************************************************
\begin{eqnarray}
\label{Eq:Sync_Eq4}
\tilde \mu_1(n,\, k)
                    & = & \tilde x_1(nT_{s1}') \tilde \beta^*(kT) \nonumber  \\
\tilde y_1(nT_{s1}')
                    & = & \sum_{i=0}^{L_p-L_{\mathrm{ISI}}-1}
                          \tilde \mu_1^*(n+iMI,\, i)              \nonumber  \\
                    &   & \mbox{ } \hspace*{-0.2in} \times
                          \tilde \mu_1(n+(i+1)MI,\, i+1).
\end{eqnarray}
%*******************************************************************************
%*******************************************************************************
\begin{figure}[tbh]
\centering
\input{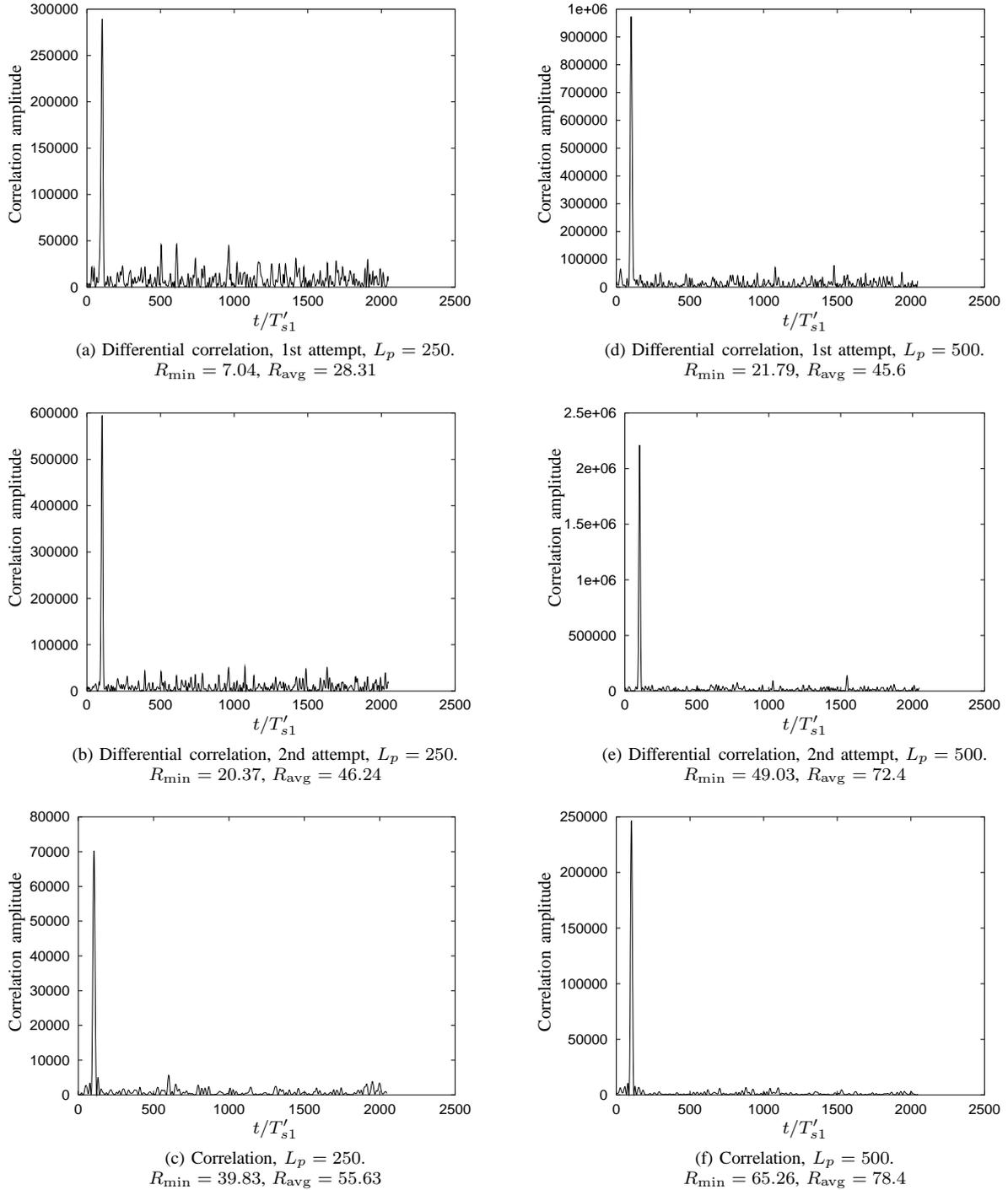}
\caption{Frame detection at 1 dB SNR for preamble lengths of $L_p=250$ and
         $L_p=500$.}
\label{Fig:Tim_Corr}
\end{figure}
%*******************************************************************************
Since the preamble and $p(t)$ are known,
$\gamma_{k,\, Q}$ and hence $\tilde \beta(kT)$ can be precomputed and stored at
the receiver for $0 \le k \le L_p-L_{\mathrm{ISI}}$ using
(\ref{Eq:Sync_Eq2}) and (\ref{Eq:Sync_Eq5_1}).

The start of frame is detected using the
following rule: choose that value of $nT_{s1}'$ which maximizes
$|\tilde y(nT_{s1}')|^2$. Mathematically, this can be stated as
\cite{Meyr98,Choi02,Vasu08}:
%*******************************************************************************
\begin{eqnarray}
\label{Eq:Sync_Eq5}
\left|
\tilde y_1(n_1T_{s1}')
\right|^2 = \max_{n}
            \quad
            \left|
            \tilde y_1(nT_{s1}')
            \right|^2.
\end{eqnarray}
%*******************************************************************************
Note that $n_1T_{s1}'$ is an estimate of $t_0$ in the first attempt.
The result of applying the detection rule in
(\ref{Eq:Sync_Eq5}) is depicted in Figure~\ref{Fig:Tim_Corr}(a) and (d)
for preamble lengths of $L_p=250$ and $L_p=500$ respectively.
Assuming that $\delta=0$, $t_0=n_1T_{s1}'$, and in the absence
of noise we have
%*******************************************************************************
\begin{eqnarray}
\label{Eq:Sync_Eq6}
\tilde y_1(n_1 T_{s1}') =
                       \mathrm{e}^{\,\mathrm{j}\, \omega_3 M}
                       \sum_{i=0}^{L_p-L_{\mathrm{ISI}}-1}
                       \left|
                       \tilde \beta(iT)
                       \tilde \beta((i+1)T)
                       \right|^2
\end{eqnarray}
%*******************************************************************************
therefore
%*******************************************************************************
\begin{eqnarray}
\label{Eq:Sync_Eq7}
\hat \omega_3 = (1/M) \arg
                      \left[
                      \tilde y_1(n_1 T_{s1}')
                      \right] \qquad \mbox{radians}.
\end{eqnarray}
%*******************************************************************************
We refer to (\ref{Eq:Sync_Eq5}) and (\ref{Eq:Sync_Eq7}) as the ``differential''
correlation method of estimating the start of frame and frequency offset.
%*******************************************************************************
\begin{table}[tbh]
\centering
\caption{Normalized variance of timing error at an SNR per bit of 1
         dB for $L_p=250$ and $\alpha=\delta=0$.}
\input{pmb250_terr.pstex_t}
\label{Tbl:PMB250_Terr}
\end{table}
%*******************************************************************************
%*******************************************************************************
\begin{table}[tbh]
\centering
\caption{Normalized variance of timing error at an SNR per bit of 1
         dB for $L_p=500$ and $\alpha=\delta=0$.}
\input{pmb500_terr.pstex_t}
\label{Tbl:PMB500_Terr}
\end{table}
%*******************************************************************************
%*******************************************************************************
\begin{table}[tbh]
\centering
\caption{RMS and maximum frequency offset estimation error in radians at an
         SNR per bit of 1 dB for $L_p=250$.}
\input{pmb250_ferr.pstex_t}
\label{Tbl:PMB250_Ferr}
\end{table}
%*******************************************************************************
%*******************************************************************************
\begin{table}[tbh]
\centering
\caption{RMS and maximum frequency offset estimation error in radians at an
         SNR per bit of 1 dB for $L_p=500$.}
\input{pmb500_ferr.pstex_t}
\label{Tbl:PMB500_Ferr}
\end{table}
%*******************************************************************************

In practice, the start of frame could be detected by computing the following
ratio for the $i^{th}$ frame:
%*******************************************************************************
\begin{eqnarray}
\label{Eq:Sync_Eq7_0}
R_i =
    \frac{\left|
    \tilde y_1(n_1T_{s1}')
    \right|^2_i}{
    \left<
    \left|
    \tilde y_1(nT_{s1}')
    \right|^2
    \right>_i}
\end{eqnarray}
%*******************************************************************************
where $<\cdot>$ denotes the time-average. In the simulations, the time average
was computed over 2048 samples (spaced at $T_{s1}'$), which includes the
peak value. It is convenient to define $R_{\mathrm{avg}}$ and
$R_{\mathrm{min}}$, which are the average and minimum values of $R_i$
over several frames. In Figure~\ref{Fig:Tim_Corr} $R_{\mathrm{min}}$ and
$R_{\mathrm{avg}}$ were computed over $10^5$ frames. We could also define
a threshold slightly less than $R_{\mathrm{min}}$. A frame could be declared as
``detected'' if $R_i$ exceeds the threshold.

The timing and frequency offset estimates obtained from
(\ref{Eq:Sync_Eq5}) and (\ref{Eq:Sync_Eq7}) are not very accurate when
$\omega_3$ is large, e.g. $0.15\pi$ radians. This is clear from
column (a) in Tables~\ref{Tbl:PMB250_Terr}, \ref{Tbl:PMB250_Ferr} for
$L_p=250$ and \ref{Tbl:PMB500_Terr}, \ref{Tbl:PMB500_Ferr} for $L_p=500$. The
main
reason is due to the approximation in (\ref{Eq:Sync_Eq1}), which gets better
as $\omega_3$ gets smaller. This is why we need to
estimate the timing and frequency offset for the second time.

The normalized (with respect to $T$) variance of the timing error in the
first attempt of the differential correlation method is given by
%*******************************************************************************
\begin{eqnarray}
\label{Eq:Sync_Eq7_1}
\left<
\left(
\frac{n_1T_{s1}'-t_0}{T}
\right)^2
\right>.
\end{eqnarray}
%*******************************************************************************
The average was computed over $10^5$ frames. The maximum timing error in
samples is given by
%*******************************************************************************
\begin{eqnarray}
\label{Eq:Sync_Eq7_1_1}
\max \frac{|n_1T_{s1}'-t_0|}{T_{s1}'}
\end{eqnarray}
%*******************************************************************************
which is taken over $10^5$ frames.
Observe that it is necessary to set $\delta=0$ and $\alpha$ equal to an
integer multiple of $T_{s1}'$ (see (\ref{Eq:Ch4_Tim_Sync_Eq7})) to compute
(\ref{Eq:Sync_Eq7_1}) and (\ref{Eq:Sync_Eq7_1_1}). In the simulations,
$\alpha$ was set to zero, for computing (\ref{Eq:Sync_Eq7_1}) and
(\ref{Eq:Sync_Eq7_1_1}).
The root mean square (rms) error in the frequency
offset estimate in the first attempt of the differential correlation
method is given by:
%*******************************************************************************
\begin{eqnarray}
\label{Eq:Sync_Eq7_2}
\sqrt{<(\omega_3-\hat\omega_3)^2>}
\end{eqnarray}
%*******************************************************************************
with the average computed over $10^5$ frames. Similarly, the maximum error
in the frequency offset estimate for the first attempt is:
%*******************************************************************************
\begin{eqnarray}
\label{Eq:Sync_Eq7_2_1}
\max |\omega_3-\hat\omega_3|
\end{eqnarray}
%*******************************************************************************

The second attempt is initiated by first demodulating $r(nT_s')$ using
the local oscillator frequency as $\pi/2 + \hat \omega_3$.
The resultant frequency offset is $\omega_f=\omega_3-\hat\omega_3$.
Let us denote the matched filter output as $\tilde x_2(nT_{s1}')$.
The effect of applying
(\ref{Eq:Sync_Eq5}) and (\ref{Eq:Sync_Eq7}) again with $\tilde x_1(nT_{s1}')$
replaced by $\tilde x_2(nT_{s1}')$, is given in
column (b) of Tables~\ref{Tbl:PMB250_Terr}, \ref{Tbl:PMB250_Ferr} for
$L_p=250$ and \ref{Tbl:PMB500_Terr}, \ref{Tbl:PMB500_Ferr} for $L_p=500$.
While there is a significant improvement in the timing estimate, the
accuracy of the frequency offset estimate is still inadequate. For example,
in Table~\ref{Tbl:PMB500_Ferr} with $L_p=500$, the second attempt yields a
root mean square (RMS) error equal to $8.8\times 10^{-3}$ radians.
With $T/T_s=M=4$, the phase change over 10
symbols is $0.0088\times 4\times 10=0.352$ radians, which is too fast for a
phase tracking loop. This motivates us to use the maximum likelihood (ML)
method of estimating $\omega_f$.

Assume that in the second attempt the outcome of
(\ref{Eq:Sync_Eq5}) with $\tilde x_1(nT_{s1}')$ replaced by
$\tilde x_2(nT_{s1}')$, is $n_2T_{s1}'$. Observe that $n_2T_{s1}'$ is the
second
estimate of $t_0$. Then (see also (\ref{Eq:Sync_Eq1}))
%*******************************************************************************
\begin{eqnarray}
\label{Eq:Sync_Eq7_2_2}
\tilde x_2(n_2T_{s1}'+kT)
                   \approx
                   \tilde \beta(kT)
                   \mathrm{e}^{\,\mathrm{j}\,(\omega_f M k+\theta_0)} +
                   \tilde v_2(n_2T_{s1}'+kT).
\end{eqnarray}
%*******************************************************************************
The ML rule for estimating the frequency
offset can be stated as follows: set $\hat \omega_f=\omega_i$ if $\omega_i$
maximizes
%*******************************************************************************
\begin{eqnarray}
\label{Eq:Sync_Eq8}
\left|
\sum_{k=0}^{L_p-L_{\mathrm{ISI}}}
\tilde x_2(n_2 T_{s1}'+kT)
\tilde \beta^*(kT)
\mathrm{e}^{-\mathrm{j}\,\omega_i  M k}
\right|
\end{eqnarray}
%*******************************************************************************
where
%*******************************************************************************
\begin{eqnarray}
\label{Eq:Sync_Eq9}
\omega_i              & = & \omega_{\mathrm{min}} + i \omega_s   \nonumber  \\
\omega_{\mathrm{min}} & < & \omega_i < \omega_{\mathrm{max}}     \nonumber  \\
\omega_{\mathrm{min}} & = & \hat \omega_3 - 0.2                  \nonumber  \\
\omega_{\mathrm{max}} & = & \hat \omega_3 + 0.2 
\end{eqnarray}
%*******************************************************************************
where $\omega_s=4\times 10^{-5}$ radians, denotes the resolution.
Observe that the FFT cannot be used
in (\ref{Eq:Sync_Eq8}), since the search is only over a narrow portion of the
digital spectrum in the range $[\hat\omega_3-0.2,\, \hat\omega_3+0.2]$. The
reason for choosing 0.2 radians can be traced to the maximum estimation
error in column (a) of Table~\ref{Tbl:PMB250_Ferr}, which is equal to
0.127 radians over $10^5$ frames. The maximum estimation error over $10^3$
frames was found (from simulations) to be 0.091 radians. Thus we find
that increase in the number of frames by two orders of magnitude, results in
only a marginal
increase in the maximum estimation error. Hence we expect the probability
of estimation error exceeding 0.2 radians, to be very small.
We now discuss the complexity of the ML approach.

For obtaining a resolution of $\omega_s=4\times 10^{-5}$ radians, the
search interval of 0.4 radians must be divided into $10^4$ frequency bins.
The complexity of the DFT is of the order of $10^4L_p$. With $L_p=250$ and
$L_p=500$, this translates to $2.5\times 10^6$ and $5\times 10^6$ operations
respectively. We now propose a two-step approach to reduce the complexity.
In the first step, we divide 0.4 radians into $B_1$ frequency bins and the
length of the DFT is taken to be $L_1<L_p$. The resolution of the first step
is thus $0.4/B_1$ radians. Let $\omega_{\mathrm{ML},\, 1}$ denote the
estimate of $\omega_f$ in the first step of the ML approach. In the second
step, the search interval is taken as
%*******************************************************************************
\begin{eqnarray}
\label{Eq:Sync_Eq9_1}
\pm 8 \times \frac{0.4}{B_1}
\end{eqnarray}
%*******************************************************************************
about $\omega_{\mathrm{ML},\, 1}$, to ensure that the maximum estimation error
lies within the search interval. This is evident from column (c) of
Tables~\ref{Tbl:PMB250_Ferr} and \ref{Tbl:PMB500_Ferr} for the parameters
given in (\ref{Eq:Sync_Eq9_2}) below. The search interval in
(\ref{Eq:Sync_Eq9_1}) is divided into $B_2$
frequency bins and the length of the DFT is $L_p-L_{\mathrm{ISI}}$, as
given in (\ref{Eq:Sync_Eq8}). The complexity of the two-step approach is
of the order of $L_1B_1+L_pB_2$. In the simulations, we have taken
%*******************************************************************************
\begin{eqnarray}
\label{Eq:Sync_Eq9_2}
L_1 & = & 100   \nonumber  \\
B_1 & = & 800   \nonumber  \\
B_2 & = & 200
\end{eqnarray}
%*******************************************************************************
so that the final resolution is equal to
%*******************************************************************************
\begin{eqnarray}
\label{Eq:Sync_Eq9_3}
16\times \frac{0.4}{B_1 B_2} = 4 \times 10^{-5} \quad \mbox{radians}
\end{eqnarray}
%*******************************************************************************
which is identical to the single stage ML approach.
With $L_p=250$ and $L_p=500$, this translates to $1.3\times 10^5$ and
$1.8\times 10^5$ operations respectively, which is more than an order of
magnitude reduction in complexity. This is shown in
Table~\ref{Tbl:ML_Complexity}. We emphasize that $L_1$, $B_1$ and $B_2$
have not been optimized to minimize the complexity. In fact, we can even have
more than two steps for reducing the complexity. This could be a subject for
future research. In any case, let $\hat\omega_f$ denote the estimate of
$\omega_f$ in the second step.

The received samples $r(nT_s')$ are demodulated again using the local
oscillator frequency equal to $\pi/2+\hat\omega_3+\hat\omega_f$.
From column (d) of Table~\ref{Tbl:PMB250_Ferr}
we find that the RMS estimation error for the ML method
is $1.52\times 10^{-4}$, for $L_p=250$. The average phase change over the
preamble is $1.52\times 10^{-4}\times 250\times 4=0.152$ radians. Similarly,
from column (d) of Table~\ref{Tbl:PMB500_Ferr}, we obtain the average
phase change over the preamble as $5.3\times 10^{-5}\times 500\times 4=0.106$
radians. These results imply that the phase can be considered to be constant
over the duration of the preamble, thereby facilitating the use of the
correlation method for estimating the start of frame for the third time. This
feature (phase being constant over a large number of symbols) also enables
the use of narrowband lowpass filters in the phase tracking loop, to
average out the effects of noise.
The symbol amplitude and the noise variance estimates (which is required for
turbo decoding) are the by-products of
the correlation method. These issues are discussed next.
%*******************************************************************************
\begin{table}[tbh]
\centering
\caption{Comparison of the complexity of the single-step and two-step ML
         method of estimating $\omega_f$. In both cases, the final resolution
         is $\omega_s=4\times 10^{-5}$ radians.}
\input{ml_complexity.pstex_t}
\label{Tbl:ML_Complexity}
\end{table}
%*******************************************************************************

Let $x_3(nT_{s1}')$ denote the samples at the matched filter output after
demodulation by $\pi/2+\hat\omega_3+\hat\omega_f$. Define
%*******************************************************************************
\begin{eqnarray}
\label{Eq:Sync_Eq9_4}
\tilde \mu_3(n,\, k)
                  & = & \tilde x_3(nT_{s1}')
                        \tilde \beta^*(kT)/
                        \left|
                        \tilde\beta(kT)
                        \right|^2                              \nonumber  \\
\tilde y_3(nT_{s1}')
                  & = & \sum_{i=0}^{L_p-L_{\mathrm{ISI}}}
                        \tilde \mu_3(n+iMI,\, i).
\end{eqnarray}
%*******************************************************************************
The correlation method for estimating the start of frame can be stated as:
%*******************************************************************************
\begin{eqnarray}
\label{Eq:Sync_Eq9_5}
\left|
\tilde y_3(n_3T_{s1}')
\right|^2 = \max_{n}
            \quad
            \left|
            \tilde y_3(nT_{s1}')
            \right|^2
\end{eqnarray}
%*******************************************************************************
where $n_3T_{s1}'$ is the third estimate of $t_0$. Observe that
%*******************************************************************************
\begin{eqnarray}
\label{Eq:Sync_Eq9_10_1}
\tilde x_3(n_3T_{s1}'+kT)
           & = & \tilde \beta(kT)
                 \mathrm{e}^{\,\mathrm{j}
                 \,((\omega_f-\hat\omega_f)Mk+\theta_0)} +
                 \tilde v_3(n_3T_{s1}'+kT)                \nonumber  \\
           & \approx
               & \tilde \beta(kT)
                 \mathrm{e}^{\,\mathrm{j}
                 \,\theta_0} +
                 \tilde v_3(n_3T_{s1}'+kT)
                 \quad \mbox{for $0\le k\le L_p-1$}
\end{eqnarray}
%*******************************************************************************
since the phase change over $L_p$ symbols due to the residual
frequency offset, $\omega_f-\hat\omega_f$, can be neglected and
$\tilde v_3(\cdot)$ denotes the noise term which has the property
\cite{Vasu_Book}:
%*******************************************************************************
\begin{eqnarray}
\label{Eq:Sync_Eq9_8_2}
(1/2) E
\left[
\tilde v_3(nT_{s1}'+kT)
\tilde v_3^*(nT_{s1}'+kT-mT)
\right] = N_0 \delta_K(mT).
\end{eqnarray}
%*******************************************************************************
Assuming that $\delta=0$, $\omega_f=\hat\omega_f$ and $t_0=n_3T_{s1}'$, and in
the absence of noise we have
%*******************************************************************************
\begin{eqnarray}
\label{Eq:Sync_Eq9_6}
\tilde y_3(n_3 T_{s1}')
                 & = & \mathrm{e}^{\,\mathrm{j}\, \theta_0}
                       \sum_{i=0}^{L_p-L_{\mathrm{ISI}}}
                       \left|
                       \tilde \beta(iT)
                       \right|^2/
                       \left|
                       \tilde \beta(iT)
                       \right|^2                     \nonumber  \\
                 & = & \mathrm{e}^{\,\mathrm{j}\, \theta_0}
                       \left(
                        L_p - L_{\mathrm{ISI}} +1
                       \right).
\end{eqnarray}
%*******************************************************************************
Therefore
%*******************************************************************************
\begin{eqnarray}
\label{Eq:Sync_Eq9_7}
\hat\theta_0 = \arg
               \left[
               \tilde y_3(n_3 T_{s1}')
               \right].
\end{eqnarray}
%*******************************************************************************
The symbol amplitude (in our case is unity) is estimated as:
%*******************************************************************************
\begin{eqnarray}
\label{Eq:Sync_Eq9_8}
\hat A = \frac{1}{L_p-L_{\mathrm{ISI}}+1}
         \Re
         \left
         \{
         \tilde y_3(n_3 T_{s1}')
         \mathrm{e}^{-\mathrm{j}\,\hat\theta_0}
         \right
         \}.
\end{eqnarray}
%*******************************************************************************
The noise variance is estimated as follows:
%*******************************************************************************
\begin{eqnarray}
\label{Eq:Sync_Eq9_9}
\hat\sigma^2      & = & \frac{1}{2(L_p-L_{\mathrm{ISI}}+1)}
                        \sum_{k=0}^{L_p-L_{\mathrm{ISI}}}
                        \left(
                        \Re
                        \left
                        \{
                        \tilde x_3(n_3T_{s1}'+kT)
                        \mathrm{e}^{-\mathrm{j}\,\hat\theta_0}
                        \right
                        \} S_{k,\, I} - \hat A
                        \right)^2                            \nonumber  \\
                  &   & \mbox{ } +
                        \left(
                        \Im
                        \left
                        \{
                        \tilde x_3(n_3T_{s1}'+kT+T/2)
                        \mathrm{e}^{-\mathrm{j}\,\hat\theta_0}
                        \right
                        \} S_{k,\, Q} - \hat A
                        \right)^2.
\end{eqnarray}
%*******************************************************************************
Note that
%*******************************************************************************
\begin{eqnarray}
\label{Eq:Sync_Eq9_10}
\tilde x_3(n_3T_{s1}'+kT+T/2)
           & = & \left(
                 \gamma_{k,\, I} + \mathrm{j}\, S_{k,\, Q}
                 \right)
                 \mathrm{e}^{\,\mathrm{j}
                 \,((\omega_f-\hat\omega_f)(Mk+M/2)+\theta_0)} +
                 \tilde v_3(n_3T_{s1}'+kT+T/2)            \nonumber  \\
           & \approx
               & \tilde\beta(kT+T/2)
                 \mathrm{e}^{\,\mathrm{j}
                 \,\theta_0} +
                 \tilde v_3(n_3T_{s1}'+kT+T/2)
                 \quad \mbox{for $0\le k\le L_p-1$}
\end{eqnarray}
%*******************************************************************************
where $\gamma_{k,\, I}$ denotes the ISI in the in-phase arm. The mean and
variance of the amplitude and noise variance estimates are discussed in the
appendix.

This completes the acquisition of the carrier frequency, symbol timing and
estimation of the symbol amplitude, carrier phase and noise variance using the
preamble. In the next subsection, we discuss the algorithms for decision
directed tracking of the carrier phase and timing, along with data detection.

%*******************************************************************************
\subsection{Decision Directed Tracking of the Timing and Carrier Phase}
\label{SSec:Sync_Track}
%*******************************************************************************
%*******************************************************************************
\begin{figure}[tbh]
\centering
\input{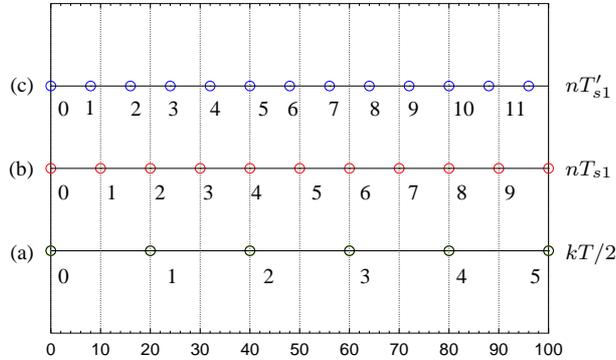}
\caption{(a) Offset QPSK symbols occur at times $kT/2$. (b) Samples taken
         at an interval of $T_{s1}=T/4$. Observe that the in-phase and
         quadrature symbols are
         obtained at times 0 and 2-mod 4 respectively. (c) Samples
         taken at an interval of $T_{s1}'=0.8T_{s1}$. The receiver does not
         know $T_{s1}'$, and it assumes that its sampling period is
         $T_{s1}=T/4$, that is, 4 samples per symbol.
         However, now the 1st in-phase symbol is obtained
         at time 0-mod 4, the 1st quadrature symbol at time 2-mod 4, the
         2nd in-phase symbol at time 1-mod 4 and so on. Therefore, the
         sampling instant changes mod-4. In any case, note that the
         symbols are obtained at times approximately equal to $kT/2$,
         independent of the sampling interval.}
\label{Fig:Tim_Err}
\end{figure}
%*******************************************************************************
Based on (\ref{Eq:Pap7_Eq10}), the sampling interval at the matched filter
output is:
%*******************************************************************************
\begin{eqnarray}
\label{Eq:Track_Eq1}
T_{s1}' = 1/F_{s1}' = T_{s1}(1 \mp 2 \delta \times 10^{-6})
\stackrel{\Delta}{=} T_{s1} + \epsilon'.
\end{eqnarray}
%*******************************************************************************
Since
%*******************************************************************************
\begin{eqnarray}
\label{Eq:Track_Eq2}
nT_{s1}' = nT_{s1} + n\epsilon'
\end{eqnarray}
%*******************************************************************************
a sample is gained or lost (in other words, the sampling instant changes) at
the matched filter output when
%*******************************************************************************
\begin{eqnarray}
\label{Eq:Track_Eq3}
n\epsilon' & \ge & T_{s1}                     \nonumber  \\
\Rightarrow
n          & \ge & 10^6/(2\delta) \quad \mbox{samples}.
\end{eqnarray}
%*******************************************************************************
The term ``sampling instant'' refers to the time when a symbol is recovered
at the matched filter output.
This is illustrated in Figure~\ref{Fig:Tim_Err}(c) for $\epsilon'=-0.2T_{s1}$.
In a more realistic situation where  $\delta=25$ ppm, we get
%*******************************************************************************
\begin{eqnarray}
\label{Eq:Track_Eq4}
n \ge 20000 \quad \mbox{samples} \equiv 20000/(MI) \quad \mbox{symbols}
=1250 \quad \mbox{symbols}
\end{eqnarray}
%*******************************************************************************
for $M=I=4$. Thus, a sample is gained or lost (the sampling instant-mod MI
changes) every 1250 symbols (recall that there are nominally $MI$ samples per
symbol at the matched filter output). Since the preamble length is only
$L_p=250$ or $L_p=500$, there is no change in the sampling instant during the
preamble. Therefore, there is no need to track the timing during the
preamble. Similarly, there is no need to track the carrier phase during the
preamble, since the residual frequency offset, $\omega_f-\hat\omega_f$ is
small enough, such that the carrier phase can be considered to be nearly
constant (refer to column (d) of Tables~\ref{Tbl:PMB250_Ferr},
\ref{Tbl:PMB500_Ferr} and (\ref{Eq:Sync_Eq9_10_1}), (\ref{Eq:Sync_Eq9_10})).
However, the data portion $L_d$ is much larger than $L_p$ (in our case
ten times), hence both timing and carrier need to be tracked. This is the
subject of this subsection.

Let
%*******************************************************************************
\begin{eqnarray}
\label{Eq:Track_Eq5}
t_1 = t_0 + L_pT \approx n_3 T_{s1}+L_p T \approx n_3 T_{s1}'+L_pT
\end{eqnarray}
%*******************************************************************************
where $t_0$ is defined in (\ref{Eq:Sync_Eq0}) and $n_3T_{s1}'$ is defined in
(\ref{Eq:Sync_Eq9_5}). Then, the first (in-phase) data
symbol is obtained at time $t_1$. Let
%*******************************************************************************
\begin{eqnarray}
\label{Eq:Track_Eq6}
\omega_r =\omega_f - \hat\omega_f.
\end{eqnarray}
%*******************************************************************************
The output of the sampler can be expressed as:
%*******************************************************************************
\begin{eqnarray}
\label{Eq:Track_Eq6_1}
\tilde x_3(t_0+kT/2) = \tilde \beta(kT/2)
                       \mathrm{e}^{\mathrm{j}\,\theta(kT/2)} +
                       \tilde v_3(t_0+kT/2)
                       \qquad \mbox{for $0\le k \le 2(L_p+L_d)-1$}
\end{eqnarray}
%*******************************************************************************
where it is understood that the in-phase and quadrature symbols are
detected when $k$ is even and odd respectively and
%*******************************************************************************
\begin{eqnarray}
\label{Eq:Sync_Eq10_1}
\theta(kT/2) = \omega_r Mk/2+\theta_0
               \qquad \mbox{for $0 \le k \le 2(L_p+L_d)-1$}.
\end{eqnarray}
%*******************************************************************************
The data-aided phase tracking loop operates at symbol-rate ($k=2l$)
as follows \cite{Vasu08}:
%*******************************************************************************
\begin{eqnarray}
\label{Eq:Sync_Eq12}
\tilde z(t_1+lT) & = &
                   \tilde x_3(t_1+lT)
                   \left(
                   \hat \beta(L_pT+lT)
                   \right)^* \bigg /
                   \left|
                   \hat \beta(L_pT+lT)
                   \right|^2                           \nonumber  \\
\tilde z_{\mathrm{avg}}(t_1+lT)
             & = & \rho_c
                   \tilde z_{\mathrm{avg}}(t_1+(l-1)T) +
                   (1-\rho_c)
                   \tilde z(t_1+lT)                    \nonumber  \\
\hat \theta(lT)
             & = & \arg
                   \left[
                   \tilde z_{\mathrm{avg}}(t_1+lT)
                   \right]
                   \qquad \mbox{for $-L_{\mathrm{ISI}}+1 \le l \le L_d-1$}
\end{eqnarray}
%*******************************************************************************
where $\hat\beta(\cdot)$ is the estimate of $\tilde\beta(\cdot)$.
The parameter $\rho_c$ was taken to be 0.97 for $L_p=250$ and 0.98 for
$L_p=500$. Note that $\hat\theta(lT)$ in
(\ref{Eq:Sync_Eq12}) is computed in the range $[0,\, 2\pi)$, hence there
is no phase ambiguity.

Observe that $\tilde z_{\mathrm{avg}}(\cdot)$ in (\ref{Eq:Sync_Eq12}) is
computed recursively. Its initial value is set to zero at the beginning of
the preamble. During the preamble, it is recursively computed at symbol-rate
as follows:
%*******************************************************************************
\begin{eqnarray}
\label{Eq:Sync_Eq12_1}
\tilde z(t_0+lT)
             & = & \tilde x_3(t_0+lT)
                   \left(
                   \tilde \beta(lT)
                   \right)^* \bigg /
                   \left|
                   \tilde \beta(lT)
                   \right|^2                               \nonumber  \\
\tilde z_{\mathrm{avg}}(t_0+lT)
             & = & \rho_c
                   \tilde z_{\mathrm{avg}}(t_0+(l-1)T) +
                   (1-\rho_c)
                   \tilde z(t_0+lT)
                   \qquad \mbox{for $0\le l \le L_p-L_{\mathrm{ISI}}$}
                                                           \nonumber  \\
\hat \theta(lT)
             & = & \arg
                   \left[
                   \tilde z_{\mathrm{avg}}(lT)
                   \right] \qquad
                   \mbox{for $l=L_p-L_{\mathrm{ISI}}$}.
\end{eqnarray}
%*******************************************************************************
After the preamble, $\tilde z_{\mathrm{avg}}(\cdot)$ is used in
(\ref{Eq:Sync_Eq12}).

Finally, the QPSK symbols at time $i=l+L_{\mathrm{ISI}}$ are estimated as
follows (note that $\tilde \beta(\cdot)$ in (\ref{Eq:Sync_Eq12}) is a function
of $L_{\mathrm{ISI}}-1$ future symbols, as given in (\ref{Eq:Sync_Eq5_1})
and Figure~\ref{Fig:ISI},
hence we assume that these symbols have already been estimated):
%*******************************************************************************
\begin{eqnarray}
\label{Eq:Sync_Eq13}
\hat S_{i,\, I} & = &
                 \mathrm{sgn}\,
                 \left[
                 \Re
                 \left
                 \{
                 \tilde x_3(t_1+iT)
                 \mathrm{e}^{-\mathrm{j}\, \hat \theta(lT)}
                 \right
                 \}
                 \right]                                    \nonumber  \\
\hat S_{i,\, Q} & = &
                 \mathrm{sgn}\,
                 \left[
                 \Im
                 \left
                 \{
                 \tilde x_3(t_1+iT+T/2)
                 \mathrm{e}^{-\mathrm{j}\, \hat \theta(lT)}
                 \right
                 \}
                 \right]        \qquad \mbox{for $0 \le i \le L_d-1$}
\end{eqnarray}
%*******************************************************************************
where $\mathrm{sgn}\,[\cdot]$ denotes the signum function.

Similarly the algorithm which tracks the timing, operates at symbol-rate, and
is initiated during the preamble. This algorithm tracks the autocorrelation
peak, $R_{pp}(0)$, which may lie in-between consecutive samples of $nT_{s1}'$.
We denote the sampling instant for obtaining the
in-phase part of a symbol as $lT$ (see Figure~\ref{Fig:Tim_Err}(c)). Define
for $0 \le l \le L_p-1$
%*******************************************************************************
\begin{eqnarray}
\label{Eq:Sync_Eq14}
u(lT+iT_{s1}')
         =    \Re
              \left
              \{
              \tilde x_3(t_0+lT+iT_{s1}')
              \mathrm{e}^{-\mathrm{j}\,\hat\theta_0}
              \right
              \}S_{l,\, I}
              \qquad \mbox{for $-2 \le i \le 2$}
\end{eqnarray}
%*******************************************************************************
where $i=0$ denotes the ``middle'' sample (the correct sampling
instant), $i< 0$ denotes the ``early'' samples and $i> 0$
denotes the ``late'' samples. Compute for $0\le l \le L_p-1$
%*******************************************************************************
\begin{eqnarray}
\label{Eq:Sync_Eq15}
u_{\mathrm{avg}}(lT+iT_{s1}')
             & = & \rho_t
                    u_{\mathrm{avg}}((l-1)T+iT_{s1}') +
                   (1-\rho_t)
                    u(lT+iT_{s1}')
                   \qquad \mbox{for $-2 \le i \le 2$}.
\end{eqnarray}
%*******************************************************************************
The initial value of $u_{\mathrm{avg}}(\cdot)$ is set to zero.
For $L_p \le l \le L_p+L_d-1$ we have
%*******************************************************************************
\begin{eqnarray}
\label{Eq:Sync_Eq16}
u(lT+iT_{s1}') =
              \Re
              \left
              \{
              \tilde x_3(t_0+lT+iT_{s1}')
              \mathrm{e}^{-\mathrm{j}\,\hat\theta((l-L_{\mathrm{ISI}})T)}
              \right
              \}
              \hat S_{l,\, I}
              \qquad \mbox{for $-2 \le i \le 2$}
\end{eqnarray}
%*******************************************************************************
where $\hat S_{l,\, I}$ obtained from
(\ref{Eq:Sync_Eq13}). Note that the averaging in (\ref{Eq:Sync_Eq15}) is done
to reduce the effects of noise and ISI.
%*******************************************************************************
\begin{figure}[tbh]
\centering
\input{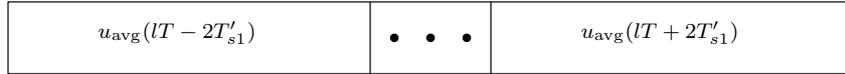}
\caption{The array elements are labeled $u_{\mathrm{avg}}(lT+iT_{s1}')$
         $-2\le i \le 2$.}
\label{Fig:Tim_Array}
\end{figure}
%*******************************************************************************

As mentioned earlier, timing is not tracked during the preamble, it is only
acquired. Tracking is done during the data part as given below. We assume
that the values of $u_{\mathrm{avg}}(lT+iT_{s1}')$ are stored in an array as
shown in Figure~\ref{Fig:Tim_Array}.
%*******************************************************************************
\begin{enumerate}
 \item For the current sampling instant $\approx lT$, for
       $L_p \le l \le L_p+L_d-1$,
       find the maximum of $u_{\mathrm{avg}}(lT+iT_{s1}')$ for
       $-1 \le i \le 1$ (not $-2 \le i \le 2$), since we expect the sampling
       instant to change by only one sample at a time.
 \item If $u_{\mathrm{avg}}(lT+iT_{s1}')$ is maximum, set $c=i$, for
       $-1 \le i \le 1$
 \item Obtain the quadrature symbol at time $lT+MIT_{s1}'/2\approx lT+T/2$
       (after $MI/2$ samples)
 \item Do the following:
%*******************************************************************************
\begin{enumerate}
 \item If $c=1$, left-shift the contents of the array by one place and
       initialize $u_{\mathrm{avg}}(lT+2T_{s1}')=0$.
 \item Else if $c=-1$, right-shift the contents of the array by one place and
       initialize $u_{\mathrm{avg}}(lT-2T_{s1}')=0$.
\end{enumerate}
%*******************************************************************************
 \item The next in-phase symbol is obtained at time
       $lT+MIT_{s1}'+cT_{s1}'\approx (k+1)T$ (after $MI+c$ samples)
 \item Reset $c=0$ and go back to step (1) with $l=l+1$
\end{enumerate}
%*******************************************************************************
Finally, the noisy symbols are obtained from (\ref{Eq:Sync_Eq13}) with the
signum function removed, and fed to the BCJR algorithm \cite{Vasu_Book} for
turbo decoding. The details of the BCJR algorithm will not be discussed here.
Having presented the various receiver algorithms, it becomes necessary to
discuss the receiver complexity. This is done in the next subsection.

%*******************************************************************************
\subsection{Receiver Complexity}
\label{SSec:Rx_Complexity}
%*******************************************************************************
We begin by discussing
the complexity of demodulation and matched filtering. Recall that this
operation has to be done thrice. However the first two times, demodulation
needs to be done only over the preamble. During the third attempt,
demodulation needs to be done over the preamble and the data. The length of
the matched filter is
97 samples (taken at a sampling frequency of $F_{s1}$), the
interpolation factor $I=4$ and the number of samples per symbol at the
input of the receiver is nominally $M=4$. Hence the number of samples per
symbol at the matched filter output is nominally $MI=16$. For obtaining each
complex output sample $(97/4)\times 2 \approx 48$ real multiplications 
and 48 real additions are required.
Therefore for obtaining all the samples corresponding to the preamble,
$48\times 3\times L_pMI$ real multiplications and the same number of real
additions are required for the three demodulation
steps. During the third step, the complexity of demodulating the data is
$48\times L_d\times 5$ real multiplications and the same number of additions,
since only five samples per symbol are computed
(see subsection~\ref{SSec:Sync_Track}). We assume that the sine and cosine
operations are performed using a table lookup. Next, we look into the
complexity of the differential correlation method.

For computing $\mu_1$ in (\ref{Eq:Sync_Eq4}), one complex multiplication is
involved. Therefore, for computing $\tilde y_1(nT_{s1})$ in
(\ref{Eq:Sync_Eq4}), approximately $2L_p$ complex multiplications and $L_p$
complex additions are involved.
Note that until a burst is detected, the only
operations performed by the receiver are demodulation and differential
correlation. The correlation method, used to detect the start of frame for the
third time requires $L_p$ complex multiplications and the same number of
complex additions. The complexity of the ML method of frequency offset
estimation has been discussed in subsection~\ref{SSec:Sync_Acq}.
The complexity of turbo decoding can be found in \cite{Vasu_Book}.

%*******************************************************************************
\begin{figure}[tbh]
\centering
\input{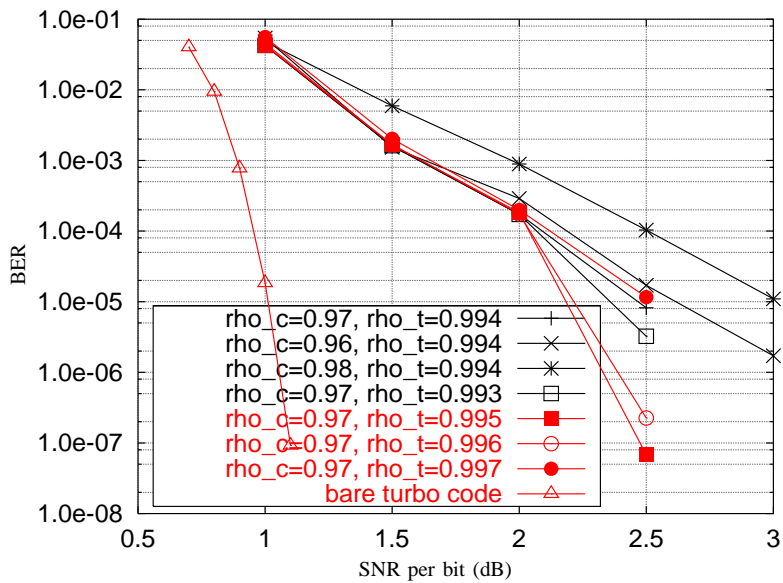}
\caption{BER performance for $L_p=250$.}
\label{Fig:BER_PMB250}
\end{figure}
%*******************************************************************************
%*******************************************************************************
\begin{figure}[tbh]
\centering
\input{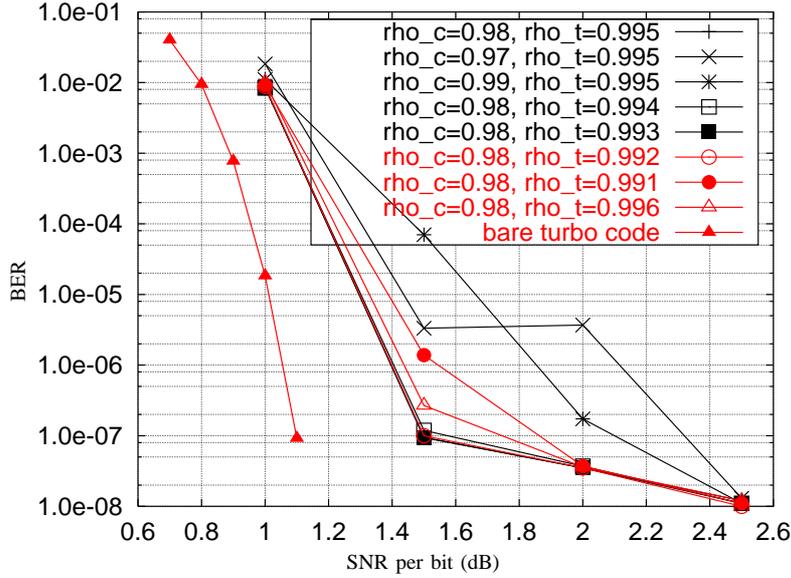}
\caption{BER performance for $L_p=500$.}
\label{Fig:BER_PMB500}
\end{figure}
%*******************************************************************************
%*******************************************************************************
\section{Simulation Results}
\label{Sec:Results}
%*******************************************************************************
In the simulations, $p(t)$ was taken to be the pulse having the root-raised
cosine spectrum with a roll-off of 0.4 and truncated to $N+1=25$ samples.
The parameter $T/T_s=M=4$ samples per symbol and interpolation factor $I=4$.
Further $L_p=250,\, 500$, $L_d=10^4$ and $L_o=12$ QPSK symbols. Simulations
were carried out over $10^5$ frames (total of $10^9$ data bits). The
frequency offset $\omega_3=0.15\pi$ (refer to the sentence after
(\ref{Eq:Pap7_Eq14})). We assume that
$\alpha$ is uniformly distributed in $[0,\, T)$ in the BER simulations. The
clock error in the transmitter and receiver is $\delta=25$ ppm, so that the
resulting offset is $2\delta=50$ ppm. The SNR per bit is defined as
\cite{Vasu_Book}:
%*******************************************************************************
\begin{eqnarray}
\label{Eq:Results_Eq0}
E_b/N_0 = 10\log_{10}
            \left(
            |S_k|^2/(2N_0)
            \right).
\end{eqnarray}
%*******************************************************************************

Finally, the BER results
are presented in Figures~\ref{Fig:BER_PMB250} and \ref{Fig:BER_PMB500} for
$L_p=250$ and $L_p=500$ respectively. For $L_p=250$, $\rho_c=0.97$ and
$\rho_t=0.995$ were found to be optimum. Similarly for $L_p=500$, $\rho_c=0.98$
and $\rho_t=0.995$ were found to be optimum. The performance loss at a BER of
$10^{-7}$ is about 1.5 dB for $L_p=250$. However, the performance loss
is only about 0.5 dB for $L_p=500$, for the same BER. For $L_p=500$ the
BER is less than $10^{-9}$ for an SNR per bit of 3 dB. Hence there is no
error floor.
%*******************************************************************************
\section{Conclusions and Future Work}
\label{Sec:Conclude}
%*******************************************************************************
In this work, we have presented discrete-time algorithms for synchronization
and detection of bursty offset QPSK signals. These algorithms can be
readily implemented on a DSP processor. The simulation parameters chosen
in this paper may not be optimum in the sense of reducing the
computational complexity without compromising the BER performance. We have
also shown via simulations that the acquisition time for a burst is equal to
the preamble length.
Future work could be in the direction of receiver design for fading channels.
%*******************************************************************************

\appendix
%*******************************************************************************
\subsection{Mean and Variance of the Amplitude and Noise Variance Estimates}
\label{SSec:Ap1_Est_Analysis}
Assuming that $t_0=n_3T_{s1}$, $\theta_0=\hat\theta_0$ in
(\ref{Eq:Sync_Eq9_8}), the mean and variance of $\hat A$ is
%*******************************************************************************
\begin{eqnarray}
\label{Eq:Sync_Eq9_8_1}
E
\left[
\hat A
\right] & = & A = 1                                       \nonumber  \\
E
\left[
\left(
\hat A-1
\right)^2
\right] & = & \frac{1}{(L_p-L_{\mathrm{ISI}}+1)^2}
               E
              \left[
              \left(
              \sum_{k=0}^{L_p-L_{\mathrm{ISI}}}
	      \frac{1}{\left|\tilde\beta(kT)\right|^2}
              \Re
              \left
              \{
              \tilde v_3(n_3 T_{s1}+kT)
              \mathrm{e}^{-\mathrm{j}\,\theta_0}
              \tilde\beta^*(kT)
              \right
              \}
              \right)^2
              \right]                                     \nonumber  \\
        & = & \frac{N_0}{(L_p-L_{\mathrm{ISI}}+1)^2}
              \sum_{k=0}^{L_p-L_{\mathrm{ISI}}}
              \frac{1}{\left|\tilde\beta(kT)\right|^2}
\end{eqnarray}
%*******************************************************************************
where we have made use of (\ref{Eq:Sync_Eq9_8_2}) and
the following relations \cite{Vasu_Book}:
%*******************************************************************************
\begin{eqnarray}
\label{Eq:Sync_Eq9_12}
\tilde v_3(\cdot)
    & \stackrel{\Delta}{=} 
        & v_{3,\, I}(\cdot) + \mathrm{j}\, v_{3,\, Q}(\cdot)   \nonumber  \\
E[v_{3,\, I}(nT_{s1}) v_{3,\, Q}(mT_{s1})]
    & = & 0 \qquad \mbox{for all $m$ and $n$}.
\end{eqnarray}
%*******************************************************************************
The rms error in the estimate of $A$ is the square root of the variance
computed in (\ref{Eq:Sync_Eq9_8_1}). The theoretical and simulated rms error
in the amplitude estimate is shown in Table~\ref{Tbl:RMS_Amp_Est_Err}.
%*******************************************************************************
\begin{table}[tbh]
\centering
\caption{The RMS error in the amplitude estimates at an
         SNR per bit of 1 dB.}
\input{rms_amp_est_err.pstex_t}
\label{Tbl:RMS_Amp_Est_Err}
\end{table}
%*******************************************************************************

Assuming that $t_0=n_3T_{s1}$, $\hat\theta_0=\theta_0$ and $\hat A = A = 1$ in
(\ref{Eq:Sync_Eq9_9}), the mean value of $\hat\sigma^2$ is
%*******************************************************************************
\begin{eqnarray}
\label{Eq:Sync_Eq9_11}
E
\left[
\hat\sigma^2
\right]  & = & \frac{1}{2(L_p-L_{\mathrm{ISI}}+1)}
               \sum_{k=0}^{L_p-L_{\mathrm{ISI}}}
                E
               \left[
               \left(
               \Re
               \left
               \{
               \tilde v_3(n_3T_{s1}+kT)
               \mathrm{e}^{-\mathrm{j}\,\theta_0}
               \right
               \} S_{k,\, I}
               \right)^2
               \right.                               \nonumber  \\
         & = & \left.              
               \mbox{ } +
               \left(
               \Im
               \left
               \{
               \tilde v_3(n_3T_{s1}+kT+T/2)
               \mathrm{e}^{-\mathrm{j}\,\theta_0}
               \right
               \} S_{k,\, Q}
               \right)^2
               \right]                               \nonumber  \\
          & = & N_0
\end{eqnarray}
%*******************************************************************************
where
%*******************************************************************************
\begin{eqnarray}
\label{Eq:Sync_Eq9_13}
S_{k,\, I}^2 = S_{k,\, Q}^2 = 1.
\end{eqnarray}
%*******************************************************************************
In order to find out the variance of the estimate of $N_0$ define:
%*******************************************************************************
\begin{eqnarray}
\label{Eq:Ap1_Eq1}
\psi_k & = & \Re
             \left
             \{
             \tilde v_3(n_3T_{s1}+kT)
             \mathrm{e}^{-\mathrm{j}\,\theta_0}
             \right
             \} S_{k,\, I}
             \qquad
             \mbox{for $0\le k \le L_p-L_{\mathrm{ISI}}$}   \nonumber  \\
\psi_{k+L_p-L_{\mathrm{ISI}}+1}
       & = & \Im
             \left
             \{
             \tilde v_3(n_3T_{s1}+kT+T/2)
             \mathrm{e}^{-\mathrm{j}\,\theta_0}
             \right
             \} S_{k,\, Q}
             \qquad
             \mbox{for $0\le k \le L_p-L_{\mathrm{ISI}}$}   \nonumber  \\
L_1    & = &  2(L_p-L_{\mathrm{ISI}}+1).
\end{eqnarray}
%*******************************************************************************
Then
%*******************************************************************************
\begin{eqnarray}
\label{Eq:Ap1_Eq2}
\hat\sigma^2 = \frac{1}{L_1}
               \sum_{k=0}^{L_1-1}
               \psi_k^2.
\end{eqnarray}
%*******************************************************************************
Note that $\psi_k$ and $\psi_i$ are uncorrelated, and being Gaussian, are also
independent. Hence
%*******************************************************************************
\begin{eqnarray}
\label{Eq:Ap1_Eq3}
E
\left[
\psi_k
\psi_i
\right] =
\left
\{
\begin{array}{cc}
 N_0 & \mbox{for $k=i$}\\
 0   & \mbox{for $k\ne i$}.
\end{array}
\right.
\end{eqnarray}
%*******************************************************************************
The variance of the estimate of $N_0$ is given by:
%*******************************************************************************
\begin{eqnarray}
\label{Eq:Ap1_Eq4}
E
\left[
\left(
\hat\sigma^2 -N_0
\right)^2
\right] & = &  E
              \left[
              \left(
              \frac{1}{L_1}
              \sum_{k=0}^{L_1-1}
              \psi_k^2 - N_0
              \right)^2
              \right]                          \nonumber  \\
        & = & \frac{1}{L_1^2}
              \sum_{k=0}^{L_1-1}
              \sum_{i=0}^{L_1-1}
               E
              \left[
              \left(
              \psi_k^2 - N_0
              \right)
              \left(
              \psi_i^2 - N_0
              \right)
              \right]                          \nonumber  \\
        & = & \frac{1}{L_1^2}
              \sum_{k=0}^{L_1-1}
               E
              \left[
              \left(
              \psi_k^2 - N_0
              \right)^2
              \right]                          \nonumber  \\
        &   & \mbox{ } +
              \frac{1}{L_1^2}
              \sum_{k=0}^{L_1-1}
              \sum_{i=0\atop i\ne k}^{L_1-1}
               E
              \left[
              \left(
              \psi_k^2 - N_0
              \right)
              \left(
              \psi_i^2 - N_0
              \right)
              \right]                          \nonumber  \\
        & = & \frac{1}{L_1^2}
              \sum_{k=0}^{L_1-1}
               E
              \left[
              \left(
              \psi_k^2 - N_0
              \right)^2
              \right]                          \nonumber  \\
        & = & \frac{2N_0^2}{L_1}.
\end{eqnarray}
%*******************************************************************************
The normalized variance of the estimate of $N_0$ can be defined as:
%*******************************************************************************
\begin{eqnarray}
\label{Eq:Ap1_Eq5}
\frac{1}{N_0}
E
\left[
\left(
\hat\sigma^2 -N_0
\right)^2
\right] = \frac{2N_0}{L_1}.
\end{eqnarray}
%*******************************************************************************
The normalized rms error in the estimate of $N_0$ is the square root of the
normalized variance computed in (\ref{Eq:Ap1_Eq5}). The theoretical and
simulated rms error in the noise variance estimate is shown in
Table~\ref{Tbl:RMS_Noise_Est_Err}.
%*******************************************************************************
\begin{table}[tbh]
\centering
\caption{The normalized RMS error in the noise variance estimates at an
         SNR per bit of 1 dB.}
\input{rms_noise_est_err.pstex_t}
\label{Tbl:RMS_Noise_Est_Err}
\end{table}
%*******************************************************************************

%*******************************************************************************
\subsection{The BCJR Algorithm \cite{Vasu_Book}}
\label{SSec:BCJR_Algo}
%*******************************************************************************
\begin{figure}[tbh]
\centering
\input{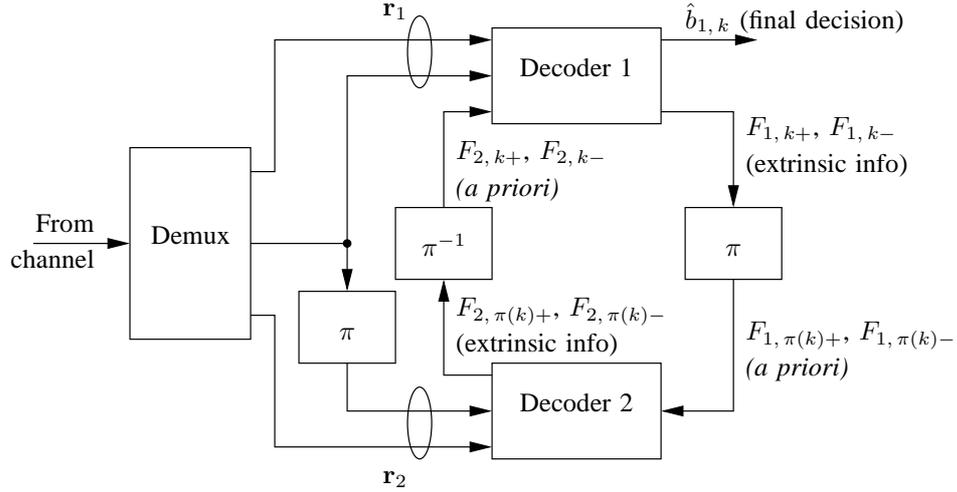}
\caption{The turbo decoder.}
\label{Fig:Turbo_Dec}
\end{figure}
%*******************************************************************************
The turbo decoder is shown in Figure~\ref{Fig:Turbo_Dec}. Assuming a code-rate
of $1/3$ and a framesize of $L$, the output of the demultiplexer is:
%*******************************************************************************
\begin{eqnarray}
\label{Eq:Turbo_Eq0_1}
r_{b1,\, m} & = & S_{b1,\, m} + w_{b1,\, m}   \nonumber  \\
r_{c1,\, m} & = & S_{c1,\, m} + w_{c1,\, m}   \nonumber  \\
r_{c2,\, m} & = & S_{c2,\, m} + w_{c2,\, m}   \nonumber  \\
r_{b2,\, m} & = & r_{b2,\, \pi(n)}
              =   r_{b1,\, n}
              =   r_{b1,\,\pi^{-1}(m)}       \qquad
                                             \mbox{for $0 \le m,\, n \le L-1$}
                                             \nonumber  \\
\end{eqnarray}
%*******************************************************************************
where $w_{b1,\, m}$, $w_{c1,\, m}$ and $w_{c2,\, m}$ are samples of zero-mean
AWGN with variance $\sigma^2_w$.
Note that all quantities in (\ref{Eq:Turbo_Eq0_1}) are real-valued. Define
%*******************************************************************************
\begin{equation}
\label{Eq:Turbo_Eq2}
\mathbf{r}_1 =
\left[
\begin{array}{lccccr}
r_{b1,\, 0} & \ldots & r_{b1,\, L-1} &
r_{c1,\, 0} & \ldots & r_{c1,\, L-1}
\end{array}
\right]^T.
\end{equation}
%*******************************************************************************
In the above equation, $r_{b1,\, k}$ and $r_{c1,\, k}$ respectively denote the
received samples corresponding to the uncoded symbol and the parity symbol
emanating from the first encoder, at time $k$. Similarly
%*******************************************************************************
\begin{eqnarray}
\label{Eq:Turbo_Eq8_1_1}
\mathbf{r}_2 =
\left[
\begin{array}{lccccr}
r_{b2,\, 0} & \ldots & r_{b2,\, L-1} &
r_{c2,\, 0} & \ldots & r_{c2,\, L-1}
\end{array}
\right].
\end{eqnarray}
%*******************************************************************************

The BCJR algorithm for turbo decoding has the following components:
%*******************************************************************************
\begin{enumerate}
 \item The forward recursion
 \item The backward recursion
 \item The computation of the extrinsic information and the final
       {\it a posteriori\/} probabilities.
\end{enumerate}
%*******************************************************************************
Let $\mathscr{S}$ denote the number of states in
the encoder trellis. Let $\mathscr{D}_n$ denote the set of states that diverge
from state $n$. For example
%*******************************************************************************
\begin{eqnarray}
\label{Eq:Turbo_Eq8_2}
\mathscr{D}_0 = \{0,\, 3\}
\end{eqnarray}
%*******************************************************************************
implies that states 0 and 3 can be reached from state 0.
Similarly, let $\mathscr{C}_n$ denote the set of states that converge
to state $n$. Let $\alpha_{i,\, n}$ denote the forward SOP at time $i$
($0 \le i \le L-2$) at state $n$ ($0 \le n \le \mathscr{S}-1$).

Then the forward SOP for decoder 1 can be recursively computed as follows
(forward recursion):
%*******************************************************************************
\begin{eqnarray}
\label{Eq:Turbo_Eq9}
\alpha_{i+1,\, n}' & = & \sum_{m \in \mathscr{C}_n}
                         \alpha_{i,\, m}
                         \gamma_{1,\,\mathrm{sys},\, i,\, m,\, n}
                         \gamma_{1,\,\mathrm{par},\, i,\, m,\, n} P
                         \left(
                          S_{b,\, i,\, m,\, n}
                         \right)                      \nonumber  \\
\alpha_{0,\, n}    & = &  1
                         \qquad
                         \mbox{for $0 \le n \le \mathscr{S}-1$}
                                                      \nonumber  \\
\alpha_{i+1,\, n}  & = & \alpha_{i+1,\, n}'\Big/
                         \left(
                         \sum_{n=0}^{\mathscr{S}-1}
                         \alpha_{i+1,\, n}'
                         \right)
\end{eqnarray}
%*******************************************************************************
where
%*******************************************************************************
\begin{eqnarray}
\label{Eq:Turbo_Eq9_1}
P(S_{b,\, i,\, m,\, n}) =
\left
\{
\begin{array}{ll}
F_{2,\, i+} & \mbox{if $S_{b,\, i,\, m,\, n}=+1$}\\
F_{2,\, i-} & \mbox{if $S_{b,\, i,\, m,\, n}=-1$}\\
\end{array}
\right.
\end{eqnarray}
%*******************************************************************************
denotes the {\it a priori\/} probability of the
systematic bit corresponding to the transition from state $m$ to state $n$,
at decoder 1 at time $i$ obtained from the $2^{nd}$ decoder at time $l$ after
deinterleaving (that is, $i=\pi^{-1}(l)$ for some $0 \le l \le L-1$,
$l\ne i$) and
%*******************************************************************************
\begin{eqnarray}
\label{Eq:Turbo_Eq10}
\gamma_{1,\,\mathrm{sys},\, i,\, m,\, n} & = &
                                              \exp
                                              \left[-
                                              \frac{
                                              \left(
                                               r_{b1,\, i}-
                                               S_{b,\, m,\, n}
                                              \right)^2}
                                              {2\sigma_w^2}
                                              \right]          \nonumber  \\
\gamma_{1,\,\mathrm{par},\, i,\, m,\, n} & = &
                                              \exp
                                              \left[-
                                              \frac{
                                              \left(
                                               r_{c1,\, i}-
                                               S_{c,\, m,\, n}
                                              \right)^2}
                                              {2\sigma_w^2}
                                              \right].
\end{eqnarray}
%*******************************************************************************
The terms $S_{b,\, m,\, n}\in\pm 1$ and $S_{c,\, m,\, n}\in\pm 1$ denote the
uncoded symbol and the
parity symbol respectively that are associated with the transition from
state $m$ to state $n$. The normalization step in the last equation of
(\ref{Eq:Turbo_Eq9}) is done to prevent numerical instabilities
\cite{Singer04}.

Similarly, let $\beta_{i,\, n}$ denote the
backward SOP at time $i$ ($1 \le i \le L-1$) at state $n$
($0 \le n \le \mathscr{S}-1$).
Then the recursion for the backward SOP (backward recursion) at decoder 1
can be written as:
%*******************************************************************************
\begin{eqnarray}
\label{Eq:Turbo_Eq11}
\beta_{i,\, n}' & = & \sum_{m \in \mathscr{D}_n}
                      \beta_{i+1,\, m}
                      \gamma_{1,\,\mathrm{sys},\, i,\, n,\, m}
                      \gamma_{1,\,\mathrm{par},\, i,\, n,\, m}P
                      \left(
                       S_{b,\, i,\, n,\, m}
                      \right)                                 \nonumber  \\
\beta_{L,\, n}  & = &  1
                      \qquad
                      \mbox{for $0 \le n \le \mathscr{S}-1$}  \nonumber  \\
\beta_{i,\, n}  & = & \beta_{i,\, n}'\Big/
                      \left(
                      \sum_{n=0}^{\mathscr{S}-1}
                      \beta_{i,\, n}'
                      \right).
\end{eqnarray}
%*******************************************************************************
Once again, the normalization step in the last equation of
(\ref{Eq:Turbo_Eq11}) is done to prevent numerical instabilities.

Let $\rho^+(n)$ denote the state that is reached from
state $n$ when the input symbol is $+1$. Similarly let $\rho^-(n)$ denote
the state that can be reached from state $n$ when the input symbol is $-1$.
Then
%*******************************************************************************
\begin{eqnarray}
\label{Eq:Turbo_Eq12}
G_{1,\,\mathrm{norm},\, k+}
       & = & \sum_{n=0}^{\mathscr{S}-1}
             \alpha_{k,\, n}
             \gamma_{1,\,\mathrm{par},\, k,\, n,\, \rho^+(n)}
             \beta_{k+1,\, \rho^+(n)}                      \nonumber  \\
G_{1,\,\mathrm{norm},\, k-}
       & = & \sum_{n=0}^{\mathscr{S}-1}
             \alpha_{k,\, n}
             \gamma_{1,\,\mathrm{par},\, k,\, n,\, \rho^-(n)}
             \beta_{k+1,\, \rho^-(n)}.
\end{eqnarray}
%*******************************************************************************
Now
%*******************************************************************************
\begin{eqnarray}
\label{Eq:Turbo_Eq12_2}
F_{1,\, k+}
& = & G_{1,\,\mathrm{norm},\, k+}/
     (G_{1,\,\mathrm{norm},\, k+} + G_{1,\,\mathrm{norm},\, k-}) \nonumber  \\
F_{1,\, k-}
& = & G_{1,\,\mathrm{norm},\, k-}/
     (G_{1,\,\mathrm{norm},\, k+} + G_{1,\,\mathrm{norm},\, k-}).
\end{eqnarray}
%*******************************************************************************
Equations (\ref{Eq:Turbo_Eq9}), (\ref{Eq:Turbo_Eq10}), (\ref{Eq:Turbo_Eq11}),
(\ref{Eq:Turbo_Eq12}) and (\ref{Eq:Turbo_Eq12_2}) constitute the MAP
recursions for the first decoder. The MAP recursions for the second decoder
are similar.

After several iterations, the final decision regarding the $k^{th}$
information bit obtained at the output of the $1^{st}$ decoder is computed as:
%*******************************************************************************
\begin{eqnarray}
\label{Eq:Turbo_Eq13}
P
\left(
S_{b1,\, k}= +1|\mathbf{r}_1
\right) & = & \sum_{n=0}^{\mathscr{S}-1}
              \alpha_{k,\, n}
              \gamma_{1,\,\mathrm{par},\, k,\, n,\, \rho^+(n)}
              \gamma_{1,\,\mathrm{sys},\, k,\, n,\, \rho^+(n)}
               F_{2,\, k+}
              \,
              \beta_{k+1,\, \rho^+(n)}                      \nonumber  \\
        & = &  F_{1,\, k+}F_{2,\, k+}
              \exp
              \left(
               -
              \frac{(r_{b1,\, k}-1)^2}{2\sigma^2_w}
              \right)                                       \nonumber  \\
P
\left(
S_{b1,\, k}= -1|\mathbf{r}_1
\right) & = & \sum_{n=0}^{\mathscr{S}-1}
              \alpha_{k,\, n}
              \gamma_{1,\,\mathrm{par},\, k,\, n,\, \rho^-(n)}
              \gamma_{1,\,\mathrm{sys},\, k,\, n,\, \rho^-(n)}
               F_{2,\, k-}
              \,
              \beta_{k+1,\, \rho^-(n)}                      \nonumber  \\
        & = &  F_{1,\, k-}F_{2,\, k-}
              \exp
              \left(
               -
              \frac{(r_{b1,\, k}+1)^2}{2\sigma^2_w}
              \right)
\end{eqnarray}
%*******************************************************************************
where again $F_{2,\, k+}$ and $F_{2,\, k-}$ denote the {\it a priori\/}
probabilities
obtained at the output of the $2^{nd}$ decoder (after deinterleaving) in the
previous iteration.

We have so far discussed the BCJR algorithm for a rate-$1/3$ encoder. In the
case of a rate-$1/2$ encoder, the following changes need to be incorporated
in the BCJR algorithm (we assume that $c_{1,\, i}$ is not transmitted for
$i=2k$ and $c_{2,\, i}$ is not transmitted for $i=2k+1$):
%*******************************************************************************
\begin{eqnarray}
\label{Eq:Turbo_Eq14}
\gamma_{1,\,\mathrm{par},\, i,\, m,\, n} & = &
                                              \left
                                              \{
                                              \begin{array}{ll}
                                              \exp
                                              \left[-
                                              \frac{
                                              \displaystyle
                                              \left(
                                               r_{c1,\, i}-
                                               S_{c,\, m,\, n}
                                              \right)^2}
                                              {\displaystyle 2\sigma_w^2}
                                              \right] & \mbox{for $i=2k+1$}\\
                                               1      & \mbox{for $i=2k$}
                                              \end{array}
                                              \right.
                                              \nonumber  \\
\gamma_{2,\,\mathrm{par},\, i,\, m,\, n} & = &
                                              \left
                                              \{
                                              \begin{array}{ll}
                                              \exp
                                              \left[-
                                              \frac{
                                              \displaystyle
                                              \left(
                                               r_{c2,\, i}-
                                               S_{c,\, m,\, n}
                                              \right)^2}
                                              {\displaystyle 2\sigma_w^2}
                                              \right] & \mbox{for $i=2k$}\\
                                               1      & \mbox{for $i=2k+1$}.
                                              \end{array}
                                              \right.
\end{eqnarray}
%*******************************************************************************

%*******************************************************************************
\bibliographystyle{IEEEtran}
\bibliography{/home/vasu/vasu/bib/mybib,/home/vasu/vasu/bib/mybib1,/home/vasu/vasu/bib/mybib2,/home/vasu/vasu/bib/mybib3,/home/vasu/vasu/bib/mybib4.bib}
%*******************************************************************************
\begin{center}
{\large \bf Publications from this Work}
\end{center}
%*******************************************************************************
\begin{itemize}
 \item K. Vasudevan, ``Synchronization of Bursty Offset QPSK Signals in the
       Presence of Frequency Offset and Noise'', Proc. IEEE TENCON, Nov. 2008.
 \item K. Vasudevan, ``Iterative Detection of Turbo-coded Offset QPSK in the
       presence of Frequency and Clock Offsets and AWGN'', Signal, Image and
       Video Processing, vol. 6, no. 4, pp. 557-—567, Nov. 2012.
\end{itemize}
%*******************************************************************************
%*******************************************************************************
\begin{center}
{\large \bf Masters Thesis Guided}
\end{center}
%*******************************************************************************
%*******************************************************************************
\begin{itemize}
 \item Deshmukh Nikhil Ashok, ``Synchronization of QPSK Signals in the
       Presence of Frequency Offset and AWGN''.
 \item Yugal Kishor Sahu, ``Application of Turbo Codes for Geosynchronous
       SATCOM Channels''.
\end{itemize}
%*******************************************************************************

\end{document}